\documentclass[preprint]{aastex}

\def\spose#1{\hbox to 0pt{#1\hss}}
\def\simlt{\mathrel{\spose{\lower 3pt\hbox{$\mathchar"218$}}
     \raise 2.0pt\hbox{$\mathchar"13C$}}}
\def\simgt{\mathrel{\spose{\lower 3pt\hbox{$\mathchar"218$}}
     \raise 2.0pt\hbox{$\mathchar"13E$}}}
\def\lsim{\rlap{$<$}{\lower 1.0ex\hbox{$\sim$}}}
\def\gsim{\rlap{$>$}{\lower 1.0ex\hbox{$\sim$}}}

\newcommand{\mum}{$\mu$m}

\usepackage[T1]{fontenc}

\usepackage{float}

\usepackage{gensymb}

\usepackage{upgreek}

\usepackage{MnSymbol}

\usepackage[usenames,dvipsnames]{color}

\usepackage{graphicx}

\usepackage{subfigure}

\usepackage{natbib}

\begin{document}

\newcommand{\fdiskstar}{F$_{disk}$/F$_{star}$}

\title{Panchromatic Imaging of a Transitional Disk: 
The Disk of GM~Aur in Optical and FUV Scattered Light}

\shorttitle{Panchromatic Imaging of a Transitional Disk}

\author{J.~B.~Hornbeck\altaffilmark{1}, 
\email{jeremy.hornbeck@louisville.edu}
J.~R.~Swearingen\altaffilmark{2},
C.~A.~Grady\altaffilmark{3,4},
\email{carol.a.grady@nasa.gov}
G.~M.~Williger\altaffilmark{1,5,6,7},
\email{gmwill06@louisville.edu}
A.~Brown\altaffilmark{8}, 
M.~L.~Sitko\altaffilmark{9,2},  
J.~P.~Wisniewski\altaffilmark{10}, 
M.~D.~Perrin\altaffilmark{11}, 
J.~T.~Lauroesch\altaffilmark{1}, 
G.~Schneider\altaffilmark{11}, 
D.~Apai\altaffilmark{12,13},  
S.~Brittain\altaffilmark{14}, 
J.~M.~Brown\altaffilmark{15},  
E.~H.~Champney\altaffilmark{2}, 
K.~Hamaguchi\altaffilmark{16,17}, 
Th.~Henning\altaffilmark{18}, 
D.~K.~Lynch\altaffilmark{19}, 
R.~Petre\altaffilmark{20}, 
R.~W.~Russell\altaffilmark{19}, 
F.~M.~Walter\altaffilmark{21}, 
B.~Woodgate\altaffilmark{20,22}}

\altaffiltext{1}{       Department of Physics \& Astronomy, University of Louisville, Louisville KY 40292 USA}
\altaffiltext{2}{       Department of Physics, University of Cincinnati, 400 Geology/Physics Bldg. PO Box 210011, Cincinnati, OH 45221-0377}
\altaffiltext{3}{       Eureka Scientific, 2452 Delmer St. Suite 100, Oakland CA 96402, USA}
\altaffiltext{4}{       Exoplanets and Stellar Astrophysics Laboratory, Code 667, Goddard Space Flight Center, Greenbelt MD 20771}
\altaffiltext{5}{       Laboratoire Lagrange, Universit\'e de Nice, UMR 7293, 06108 Nice Cedex 2, France}
\altaffiltext{6}{       IACS, Dept. of Physics, Catholic University of America, Washington DC 20064}
\altaffiltext{7}{       Jeremiah Horrocks Inst., U Central Lancashire, Preston PR1 2HE, UK}
\altaffiltext{8}{       CASA, University of Colorado, Boulder, CO 80309-0593, USA}
\altaffiltext{9}{       Space Science Institute, 4750 Walnut Street Suite 205, Boulder, Colorado 80301}
\altaffiltext{10}{      Homer L. Dodge Department of Physics \& Astronomy, University of Oklahoma, 440 W Brooks St, Norman, OK 73019}
\altaffiltext{11}{      Space Telescope Science Institute, 3700 San Martin Drive, Baltimore, MD 21218, USA}
\altaffiltext{12}{      Department of Astronomy and Steward Observatory, University of Arizona, 933 N. Cherry Avenue, Tucson, AZ 85721}
\altaffiltext{13}{      Department of Planetary Sciences and Lunar and Planetary Laboratory, University of Arizona, 1629 E. University Blvd., Tucson, AZ 85721-0092}
\altaffiltext{14}{      Department of Physics \& Astronomy, Clemson University, 118 Kinard Laboratory, Clemson, SC 29634-0978}
\altaffiltext{15}{      Harvard-Smithsonian Center for Astrophysics, 60 Garden Street, Cambridge, MA 02138, USA}
\altaffiltext{16}{      Department of Physics, UMBC, Baltimore, MD 21250, USA}
\altaffiltext{17}{      CRESST and X-ray Astrophysics Laboratory NASA/GSFC, Greenbelt, MD 20771}
\altaffiltext{18}{      Max Planck Institut F\"{u}r  Astronomie K\"{o}nigstuhl 17 D-69117 Heidelberg}
\altaffiltext{19}{      The Aerospace Corporation, Los Angeles, CA 90009}
\altaffiltext{20}{      NASA's Goddard Space Flight Center, Greenbelt MD 20771, USA}
\altaffiltext{21}{      Department of Physics and Astronomy, Z=3800, Stony Brook University Stony Brook NY 11794-3800}
\altaffiltext{22}{      deceased}

\begin{abstract}

We have imaged GM~Aurigae with the Hubble Space
Telescope (HST), detected its disk in scattered light 
at 1400~\AA\ and 1650~\AA, and compared these with observations at 3300~\AA, 
5550~\AA, 1.1$\mu$m, and 1.6$\mu$m.  The scattered light increases at
shorter wavelengths. The radial surface brightness profile at 3300~\AA\ shows no
evidence of the 24~AU radius cavity that has been previously observed in
sub-mm observations.    Comparison with dust grain opacity models
indicates the surface of the entire disk is populated with sub-$\mu$m grains. 
We have compiled an SED from 0.1~$\mu$m to 1~mm, and used it
to constrain a model of the star~+~disk system that includes 
the sub-mm cavity using the Monte Carlo Radiative Transfer (MCRT) 
code by Barbara Whitney.  
The best-fit model image indicates that the cavity 
should be detectable in the F330W bandpass if the cavity has been 
cleared of both large and small dust grains, but we do not detect it.
The lack of an observed cavity can be explained by 
the presence of sub-$\mu$m grains interior to the sub-mm cavity wall.  
We suggest one explanation for this which could be due to a 
planet of mass $<$~9~M$_J$ interior to 24~AU.  
A unique cylindrical structure is detected in the FUV data from the
Advanced Camera for Surveys/Solar Blind Channel
(ACS/SBC). It is  
aligned along the system semi-minor axis, but does not resemble an accretion-driven jet.  
The structure is limb-brightened and extends 190~AU~$\pm$~35~AU above the disk 
midplane.  The inner radius of the limb-brightening is 40~$\pm$~10~AU,
just beyond the sub-millimeter cavity wall.

\end{abstract}

\keywords{ 
protoplanetary disks -- 
stars: circumstellar matter -- 
stars: individual (GM~Aur)
stars: protostars -- 
stars: variables: Herbig Ae/Be -- 
ultraviolet: planetary systems
}

\section{Introduction}

Transitional disks are protoplanetary disks that are in the process of evolving from 
a gas-rich primordial disk to a gas-poor debris disk.  During this transition period, 
material in the disk within the first few tens of AU from the star clears to form 
an optically thin gap or cavity.  
Consistent with the presence of accretion, these objects retain an inner 
disk within a few AU of the star \citep{Hartigan_90}. If the cleared cavities 
were devoid of material, these systems would be unable to replenish their 
inner disk and accretion would cease relatively quickly.
Many of these objects continue to accrete
material while having cavities tens of AU in radii, as evidenced by
mid-IR/sub-mm/mm observations \citep{Rice_06,Lubow_06,Salyk_13}.  

One possible
explanation for such cavity generation
is a filtration mechanism that allows only small dust grains,
which would be
undetected in the sub-mm/mm data, to migrate inward through the gap entrained
with the still-accreting gas from the
outer disk \citep{Quillen_04}. \citet{Cieza_12} compared several processes
driving disk evoution including
grain growth, the effect
of planets and photoevaporation mechanisms. Starting with the hypothesis
that small grains in
a cavity are part of the process of grain growth,  
they found that grain growth can account for
$>40$\% of transition disks around K- and M-type stars, though the process can
have complicating factors like fragmentation or replenishment \citep{Dullemond_05}.
The dust grains that make up the interstellar medium (ISM) are estimated to
include a range in size from 0.005~$\mu$m to 1~$\mu$m \citep{Mathis_77}. 
\citet{Rice_06} suggest that filtration of small ISM-like grains into the cavity
devoid of larger grains could be an indication that a planet with a mass of
1-6~M$_J$ resides within the cavity.  Simulations by  \citet{Zhu_12} and
\citet{de_Juan_Ovelar_13} produce similar results, and they also suggest the 
presence of one or more gas giant planets as a disk clearing mechanism.    
FUV-driven neutral atomic or molecular disk winds have also been  considered as
a possible mechanism for disk dispersal.  At large radii, giant planets 
have difficulty clearing material on a sufficiently short time scale to be consistent with 
estimates of the disk-dispersal time for T~Tauri stars 
\citep[$\approx 10^5$ yr;][]{Simon_95,Wolk_96}. 
In rich cluster environments, \citet{Johnstone_98} showed that FUV radiation from nearby stars 
can dominate the photoevaporation rate. This, however, does not explain the rapid dispersal of disks 
around young stars with no nearby source of the FUV flux. \citet{Gorti_09} have considered flows 
driven by FUV radiation from the central star and argued that mass-loss rates of the 
order of 10$^{-8}$~M$_{\sun}$~yr$^{-1}$ can be obtained at large radius (>100 AU). 

Mapping the dust grain size distribution throughout the disk can indicate which 
mechanism is responsible for clearing material from the disk. Dust grains
scatter light most efficiently at wavelengths comparable to their size, so
determining whether sub-$\mu$m dust grains are present inside a disk cavity
requires observations to be in sub-$\mu$m bandpasses.  Disks with large cavities
provide test-beds for multi-wavelength observations probing the surface of the
disk to analyze the light scattering properties and size distribution of the
dust grains.  This method for determining grain properties and
distributions has been employed successfully at sub-mm and longer wavelengths
\citep{Perez_12, Banzatti_11}, but relies on scattered light  rather than
thermal emission.  High contrast far-ultraviolet (FUV) and optical images have
the added benefit of a smaller inner working angle (IWA) than at longer
wavelengths with the same instrument. 
For disks with cavities tens of AU in diameter, such observations 
 allow us to map the spatial distribution of gas and small
(sub-$\mu$m) grain reflection nebulosity present at the disk surface, 
both interior and exterior to the cavity wall.  
If small-grain dust
exists interior to the large grain cavity wall, we would not expect 
transitional disks with relatively high accretion rates to have an optically thin 
cavity at short (FUV and optical) wavelengths. 

The disk associated with the classical T~Tau star
GM~Aur \citep[K5.5 $\pm$ 1.0, B-V=1.12,][]{Espaillat_10} 
has been studied extensively for two decades with
the Hubble Space Telescope (HST).  The distance to GM~Aur 
is 136$_{-29}^{+50}$ pc
\citep{Bertout_06}, 
and the inclination from pole-on for its disk is 55\degree 
\citep{Calvet_05,Hughes_09,Andrews_11}.  The disk has been detected
in scattered optical and NIR light \citep{Stapelfeldt_95,Schneider_03}.  
The disk became categorized as transitional when millimeter and 
sub-millimeter observations detected a cleared inner cavity  extending 
from the star to between 20~AU and~28~AU \citep{Calvet_05,Hughes_09,Grafe_11,Andrews_11};  
in this paper we assume the sub-mm cavity is 24~AU in radius.   
GM~Aur is also accreting material at a rate of
$\approx$ 10$^{-8}$ M$_{\sun}$ yr$^{-1}$ \citep{Gullbring_98,White_01,Ingleby_15}, suggesting that
material is migrating inward from the outer disk, through the sub-mm cavity, 
and accreting onto the star.  This can be directly tested
with FUV and short wavelength optical observations.  We have acquired the requisite
datasets with HST's Advanced Camera for Surveys (ACS) using the Solar-Blind Channel (SBC),
as well as archival data from both the High Resolution Channel (HRC) and the 2nd
generation Wide Field Planetary Camera (WFPC2).

In this paper HST data of GM~Aur, along with data from a wide variety
of other instruments, 
are used to create a Spectral Energy
Distribution (SED), all of 
which we then use to model the GM~Aur star+disk system.
We probe material within the sub-mm cavity region, to test 
the hypothesis that small grain dust exists within the 24~AU cavity of GM~Aur's 
transitional disk.  A description of the data used and the data 
reduction are described in \S~2.  Our analysis and results
are found in \S~3, with a discussion in \S~4.

\section{Observations \& Data Reduction}

GM~Aurigae has been observed by HST at multiple epochs over wavelengths ranging
from the FUV to NIR, and at longer wavelengths by other instruments. 
The HST/NIR observations obtained by the Near Infrared Camera and Multi-Object
Spectrometer (NICMOS) were analyzed and discussed in \citet{Schneider_03}. 
The HST WFPC2 F555W, ACS/HRC F330W and ACS/SBC F140LP and F165LP data sets
have not 
previously been addressed in detail and so
are presented here. 
The central wavelengths, bandpasses (FWHMs), spatial sampling and angular
resolution from \citet{ACS_instrument_handbook} and \citet{WFPC2_instrument_handbook}
are summarized
in Table~\ref{tab-centralwavelengths}.
The longer wavelength WFPC2 data were taken with very short and 20 times
deeper exposures such that the short exposures were underexposed, while the
longer exposures were highly saturated.  
We have therefore excluded them from our analysis.  
One of our goals is to examine the surface brightness
profile as close to the edge of the sub-mm cavity as possible.
We can do this with the F555W and shorter wavelength observations. 
Our goals also include 
estimating the effects of (i) 
random errors in the scatter from multiple observations of GM~Aur and
a given PSF-template star (possible for F555W observations), 
and (ii) systematic errors arising
from the choice of one PSF-template star vs. another (done for F330W
observations).  
A detailed list of the HST observations we have used
is in Table~\ref{tab-observations}.

\subsection{WFPC2 Broadband Optical Data - F555W \label{WFPC2_555}}

GM~Aur was observed as a suite of $4\times 30$~s exposures with WFPC2  and the 
F555W filter, using a 4-pt, 20-pixel (0$\farcs$91 )  dither pattern on the PC1
chip (PI Trauger).  The WFPC2 data were processed and reduced with
the On-The-Fly Reprocessing  (OTFR) pipeline, yielding four flat-fielded
individual (undrizzled) images prior to our analysis
\citep{WFPC2_data_handbook}.  Our starting point for the data reduction was the
pipeline-processed  science image (c0m.fits) files which retain the full
sampling of the  planetary camera chip data.   The disk is detected in the raw
data at each dither position, but  the disk/star contrast can be
improved by subtraction of  suitably scaled and registered PSF template data. We
checked the MAST archive for a suitably exposed PSF template star for this
WFPC2/PC1 configuration.  Three template stars were considered: the er star
HIP~66578, the  sdF~star HIP~109558, and the G5IV star HD~283572.  The data for
HIP~66578 and HIP~109558 were obtained for photometric  calibration of WFPC2,
but were shallow exposures which do  not map the wings of the stellar point
spread function much beyond  the first Airy ring. As a result, we have made use
of the data  for HD~283572 \citep[G5IV, B-V=0.77,][]{Strassmeier_09}.
\footnote{The original reference to HD~283572 is \citet{Walter_87}, who
found a slightly redder value of $B-V=0.83$.}  
HD~283572 was observed as part of the same program (HST-GTO-6223)  and has data
with the same relative exposure depth and  dither pattern as for GM~Aur.  Both
datasets are slightly overexposed  and show some bleeding of charge in the
detector  +y direction over $\pm$2-3 pixels. The HD~283572 and GM~Aur images
were taken close in time (6 days apart), space ($7.6\degree$ apart on the sky)
and orientation ($4.3\degree$ difference in orientation  angle).   We did not
find any other satisfactory PSF star.

\begin{figure*}[htbp]
\includegraphics[scale=0.7]{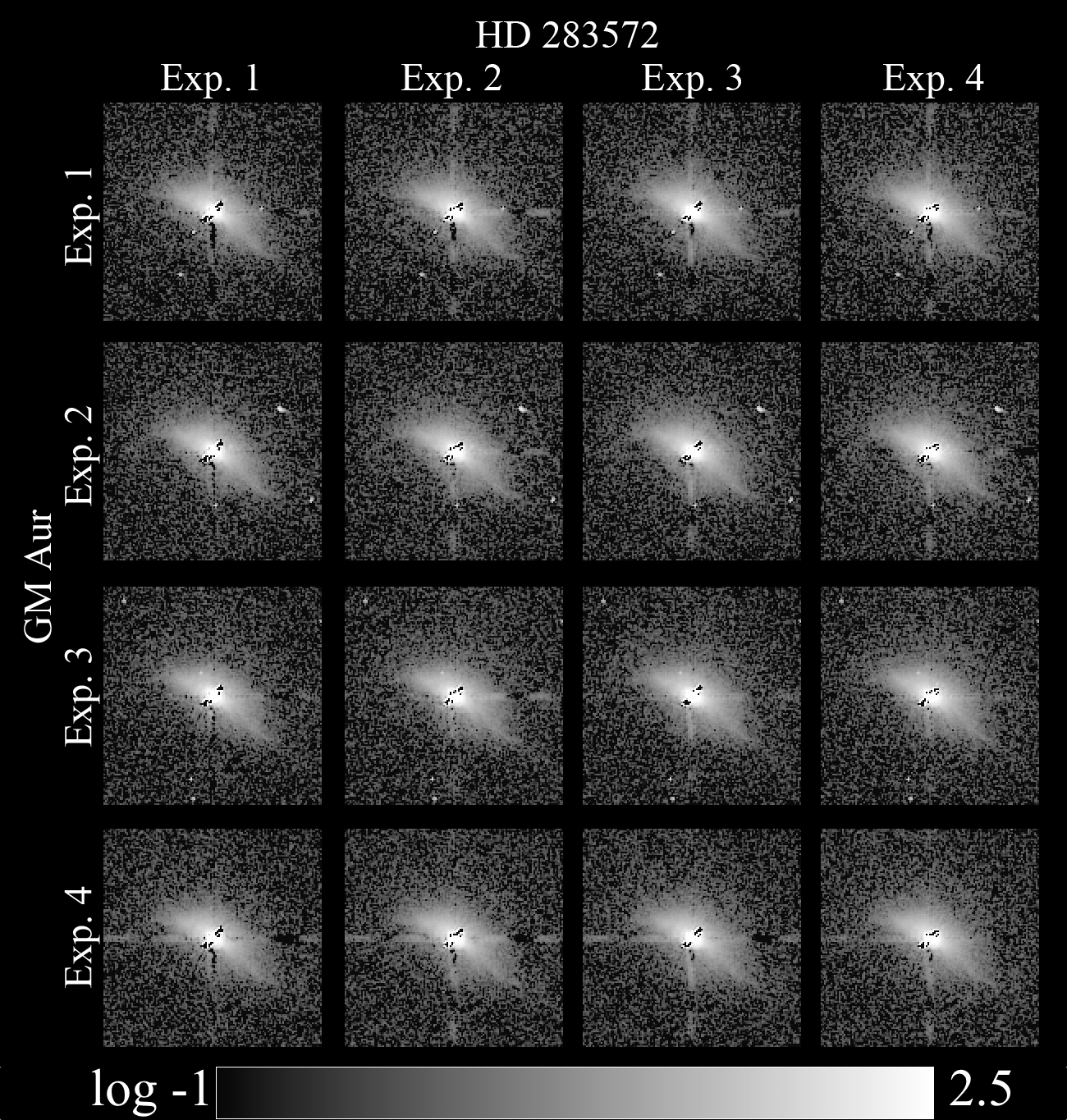}
\caption{The four exposures of GM~Aur in the F555W filter subtracted from each of the four exposures of HD~283572.  
We attribute the variation between exposures to 
environmental stresses on the spacecraft while in orbit.  Images
have been rotated to show N up and E to the left. 
The combination
of GM~Aur exposure~3 and HD~283572 exposure~4 were used for illustration
in Fig.~\ref{fig-psfsubtractions}.}
\label{fig-f555w_exposuresubtr_grid}
\end{figure*}

\begin{figure}[ht]
\centering
\includegraphics[scale=0.25]{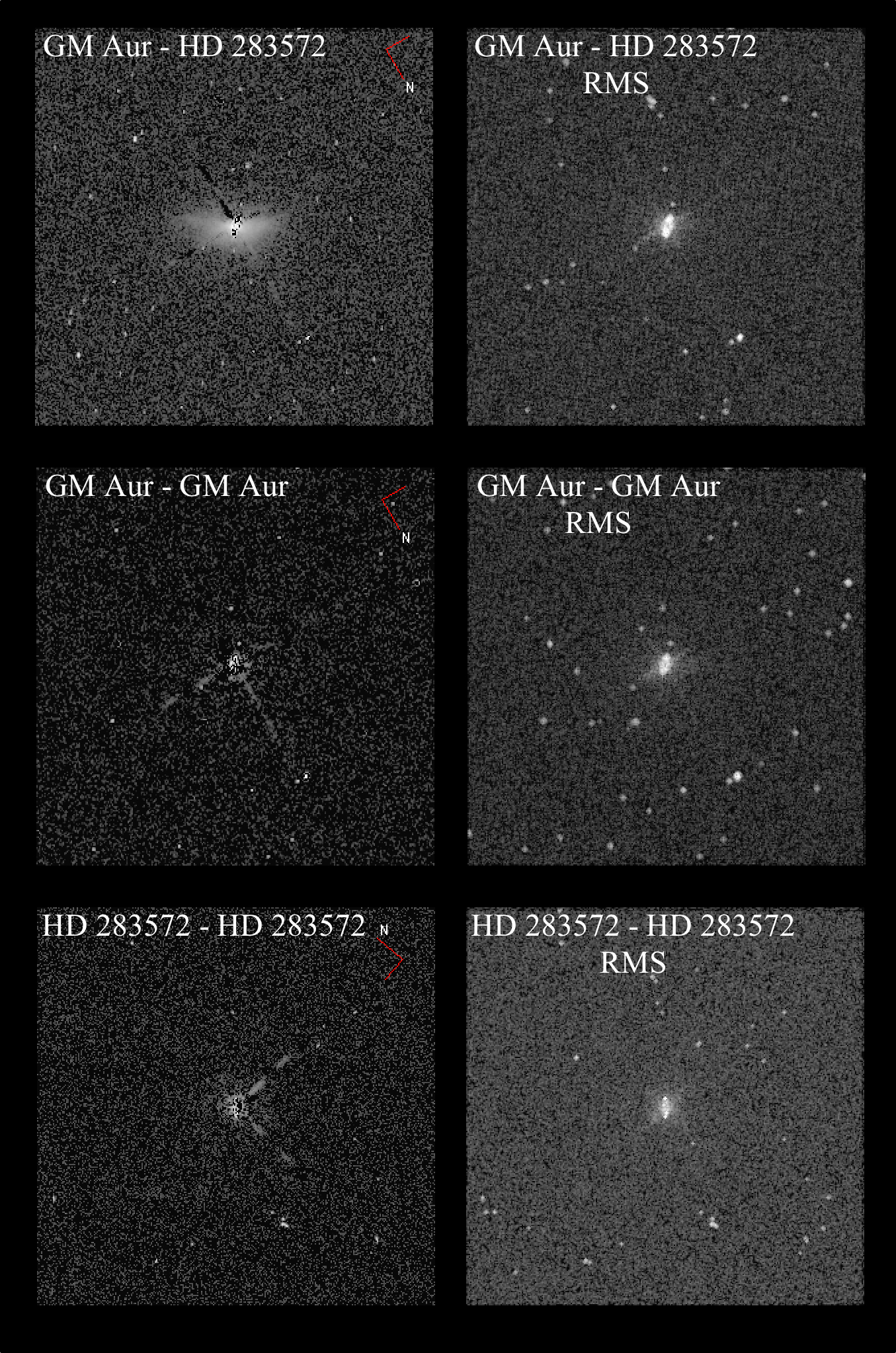}
\caption{To determine the significance of variations between individual exposures we 
compared and RMS image of the PSF-subtracted data for GM Aur - HD~283572 with 
subtractions of GM Aur from itself and HD~283572 from itself.  Unsurprisingly, the 
pixel to pixel variations are smallest for the self-subtracted images. See
the text for values.  \label{F555W_subwithRMS}}

\end{figure}

We constructed sixteen target-PSF template image pairs.  These image pairs
were  registered to the first of the GM~Aur observations, and then we 
calculated a median (to eliminate cosmic rays/bad pixels) and a  standard
deviation image using the IDL routine Image Display Paradigm \#3 ({\sc IDP3})
\citep{Lytle_99,Schneider_02}.  Manual PSF subtraction using a method nearly
identical to this was described by \citet{Krist_00}, doing the photometry by
``counting the number of saturated pixels and converting to fluxes assuming a
full-well electron depth, as described by \citet{Gilliland_94}. These values were
then multiplied by a factor of 1.10 to account for flux outside of the saturated
regions (determined by masking out the same pixels in simulated PSFs).''
\citet{Krist_00} note that their method of estimating fluxes in the case of using
saturated pixels is good to $\sim 5$\%. Diffraction spike residuals are visible
in the median image, as  expected for image sets where there are changes in HST
focus after  slews (``breathing'') and where the color match between  the PSF
template and the target is imperfect.   We then computed  median and standard
deviation images for the GM~Aur-GM~Aur data and  for PSF-PSF and find similar
residuals  and variability {\it for both stars},  consistent with detection of
stellar activity and/or accretion luminosity variations. Variation in the
residual diffraction spikes was seen between image pairs,  at a level consistent
with HST thermally-driven focus changes and stellar activity in the raw data,
for {\it both} GM~Aur and the  PSF template object HD~283572. Each of the
sixteen GM~Aur-HD~283572 image pairs is shown in
Fig.~\ref{fig-f555w_exposuresubtr_grid} to illustrate the variation between   %fig 1
exposures.  We also tested the variation between individual exposures 
by subtracting exposures of each object from a 
different exposure of itself (Fig.~\ref{F555W_subwithRMS}). 
The mean and standard deviation of the RMS values
around the disk (within 1$\farcs$2 of GM~Aur) for the GM~Aur -- PSF subtractions
are $3.1\pm 2.0$ counts sampled in four widely-separated test
regions avoiding diffraction spikes and
containing between 40 and 138 pixels each. 
For the GM~Aur -- GM~Aur
images for two different exposures the RMS values are $0.9\pm 0.1$ counts in the same
four test regions, while for a pair
of PSF - PSF images they are $1.1\pm 0.3$.  Further away, about $4''$ from the flux centroid,
the RMS values for GM~Aur -- PSF are $0.6\pm 0.1$ for four larger background
test regions containing 900-2800 pixels each.
The GM~Aur -- GM~Aur values are $0.5\pm 0.1$, while for the PSF -- PSF subtraction they are 
$0.8\pm 0.1$.  We conclude that the individual exposure variations contribute on the order
of a third of the noise level compared to a single GM~Aur frame with a single PSF
frame subtracted from it.

%fig 2

To further characterize the uncertainty arising from  individual GM~Aur and HD~283572
exposures, the standard deviation for each pixel was calculated based on each of
the 16 subtraction pairs. Simple photometry with a $13\times 7$ pixel
rectangular aperture for GM~Aur shows a flux variation of 0.3\% between the
exposures, while a $16\times 7$ rectangular aperture for HD~283572 reveals
variations of up to 2.8\%\ between exposures, with the third exposure having the
highest flux.  These variations are less than the 5\% quoted accuracy for
photometry based on counting saturated pixels \citep{Krist_00}. We frequently
could null two of the diffraction spikes satisfactorily, but could not
simultaneously null all four diffraction spikes completely for any of the 16
exposure-subtraction  combinations. The variations in nulling efficiency show a
spread of scale factors on the order of 25~-~30\%. These variations dominate
over other sources of uncertainty when  determining the disk-scattering 
fraction ($F_{disk}/F_{star}$). Although there is only one suitable
PSF template available in the HST archive for F555W, we estimate
the uncertainty from PSF template selection to be similar to that derived
for F330W, on the order of $\sim 25$\% (see below).

\subsection{HRC Optical Data - F330W}
\label{subsubsec-data-f330w}

For our analysis, we use the pipeline-processed, calibrated
flat-fielded  (flt) files for the one 360~s F330W exposure employed for GM~Aur 
(PI Hartigan).  We checked the MAST
archive for a suitably exposed star to act as a satisfactory PSF template, and
identified HIP~66578, HIP~109558, HD~202560 (AX~Mic), CY~Tau and DS~Tau as
possible  candidates for this ACS/HRC configuration.  

As with the F555W data, we scaled each of the PSF templates to optimally null
the diffraction spikes.    We varied the scale factor  so that the profile of
the residual diffraction spike was within $\le \pm\ 1\sigma$ from the local
mean, determined from local root mean square variations in the data.  This gave
the bounds for the uncertainty in the scale factor, and at the same time nulled
any tiger-striping/Airy ring residuals.  We compared each of the PSF
subtractions  from GM~Aur for each template star individually, to estimate the
(systematic)  uncertainty arising from PSF template selection. We similarly
created a $\sigma$ image for each PSF subtraction, for statistical error use
with the surface brightness profile calculation. The uncertainty in the scale
factor for the PSF subtraction is the dominant source of flux uncertainty in
each of the PSF subtracted images, on the order of $\sim 25$\%.

\begin{figure*}[htbp]
\includegraphics[scale=0.85]{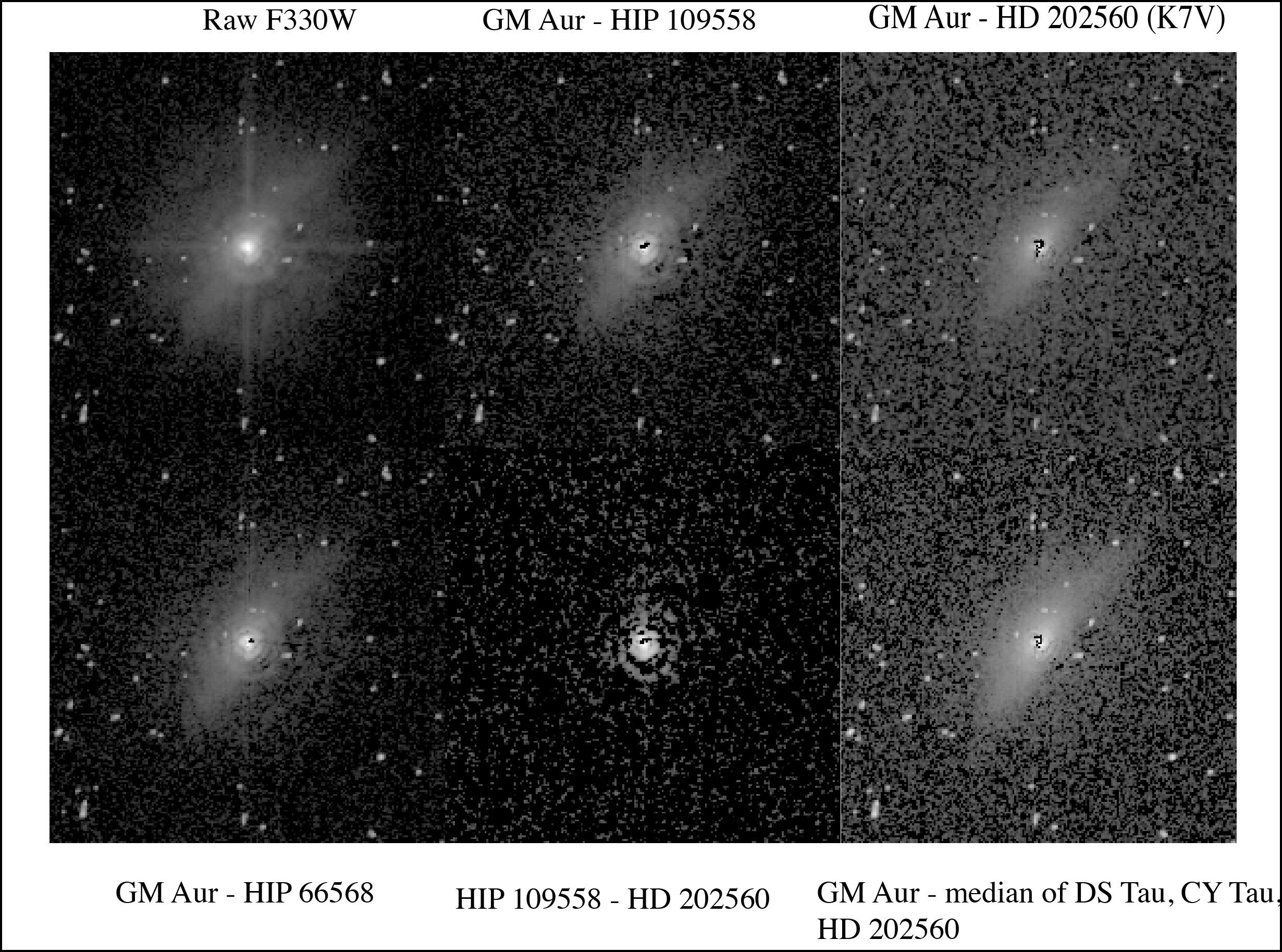}
\caption{Example subtractions for the F330W images of GM~Aur and a series of PSF standard
stars.  {\it Upper left:} raw image. {\it Lower left and upper center:}
subtractions with HIP~66578 and HIP~109558. {\it Lower center:} HIP~109558 - 
HD~202560, with the residual indicating a color difference. {\it Upper and lower
right:} subtractions with HD~202560 and a median of HD~202560, CY~Tau and DS~Tau.}
\label{fig-f330w_samplesubtractions}
\end{figure*}

The subtractions for the bluest two stars, HIP~66578 (white dwarf DA2.4) and
HIP~109558 (sdF8), have Airy ring residuals, indicating color mismatches. The
best color matches came from the nearby flare star HD~202560 
(M0V)\footnote{These spectral types are from simbad.u-strasbg.fr/simbad}, and
the classical T~Tau stars CY~Tau (M1.5) and  DS~Tau (K4V:e,
Fig.~\ref{fig-f330w_samplesubtractions}).   MAST contains a single 360~s		%fig 3
exposure for CY~Tau and DS~Tau, and $2\times 2$~s exposures for HD~202560.  We
ended up using a median of the three stars for the PSF template.  
The variation in the PSF-subtracted disk flux minus the PSF flux is
$\sim 6-8\%$ for the PSF derived from the median of the three stars, vs.
each individual PSF template.

\subsection{SBC FUV Data - F140LP \& F165LP}
\label{subsubsec-data-f140lp}

We employ GM~Aur images in the FUV with the ACS/SBC using the F140LP and F165LP
filters (PI A. Brown).   
This observing strategy was chosen to exclude
geocoronal Ly$\alpha$ and \ion{O}{1}.

We used the flatfielded, uncombined/individual
exposures (flt), uncorrected for geometric distortion, initially processed via the OTFR
pipeline \citep{ACS_data_handbook}.  

PSF subtractions are best done in the detector frame
with images that have not been corrected for geometric distortion.
The correction mapping will change between orientations, and
the PSF primarily refers to the detector frame.  
The data reduction pipeline has historically concatenated the geometric
distortions, which would be detrimental to PSF subtraction.
Additionally, the PSF subtraction
deals with scales of 1--2~arcsec, whereas the geometric distortion becomes
significant on scales approaching the order of the field of view.
We queried the STScI Help Desk in March 2016, and were
told that the last time the geometric distortion was updated was in 2008,
with no further updates currently planned.

There are not many choices available for suitable PSF templates with the
F140LP and
F165LP filters.  We considered the white dwarf
GD~71 \citep[$B-V=-0.249$,][]{Landolt_92} which is likely an extreme in color match. 
We also
examined NQ~UMa \citep[G9V,
$B-V=0.81$,][]{Montes_01},\footnote{simbad.u-strasbg.fr/simbad}.  
The $B-V$ color is typical of a G9V star, so we conclude that
there is no significant foreground extinction.
NQ~UMa
was observed in the FUV with the ACS/SBC 
(PI C. Grady) with the
F140LP and F165LP filters \citep{Hornbeck_12}.  
Of the two, NQ~UMa is the closer color match to the photosphere, 
so we consider it the preferred PSF template between the two.
The integration time in each bandpass for NQ~UMa was 2648~s.   We used the
analogous (flatfielded, individual) exposures, similar to how we worked with the
F555W data. Inspection of the raw data for GM~Aur demonstrates that the image
lacks the bright core typical of both unresolved sources (NQ~UMa or GD~71). 
Since GD~71
has a color difference with GM Aur, the effect cannot be chromatic.
In the FUV, the light from GM Aur is not dominated by an unresolved source, but
rather by extended emission.
For both GM~Aur and NQ~UMa, for 
presentation purposes, we binned the pixels slightly along the y-axis to
produce $\approx$ 0$\farcs$034 $\times$ 0$\farcs$034 pixel$^{-1}$.

The PSF subtraction process for the ACS/SBC data was the same as 
for the longer wavelength images, except that the nulling was done
on the core of GM~Aur, because no diffraction spikes were visible,
indicating that $F_{disk}/F_{star}$ was large. 
The lower bound for the scale factor was taken at the point where the
peak of the PSF standard  matched the peak of the GM~Aur central pixel 
flux.  The
upper bound was taken by requiring the flux in an inner 5 pixel radius 
to be matched for the PSF standard and GM~Aur.  We can measure flux
to an inner radius of $\approx$ 0$\farcs$1 (a three pixel radius).

\subsection{Other Data}
\label{subsec-otherdata}

Modeling the SED of GM~Aur over the widest possible wavelength range required 
additional data from a series of data archives and ground-based telescopes.  
Fortunately,
a thorough collection of data across a large wavelength range exists for 
this purpose.
The data we use provide wavelength coverage from 0.1~$\mu$m to 1000~$\mu$m. 
The FUV, NUV, and optical spectroscopic data were obtained by STIS on 2011-09-11. Optical and NIR 
spectrometry was obtained with SpeX at the NASA Infrared Telescope Facility (IRTF)
from 0.7~$\mu$m to 5.3~$\mu$m over 5 nights in 2011 September and 2012 January \citep{Ingleby_15}. 
The NIR photometric data points for GM~Aur were obtained with the Two 
Micron All Sky Survey (2MASS). Our near to mid-IR data for GM~Aur come 
from the Wide-Field Infrared Survey Explorer (WISE),  Spitzer Infrared 
Spectrograph (IRS), and AKARI/IRC data \citep{AKARI_IRC}. Our mid to far-IR 
data points were obtained with AKARI-FIS and IRAS, and additional data supplied 
by the following literature: AEF90 = \citet{Adams_90}, AW05 = \citet{Andrews_05}, 
WSD89 = \citet{Weintraub_89}, and KSB93 = \citet{Koerner_93}.  

\section{Results and Analysis}

\subsection{High Contrast Imagery from HST}
\label{subsec-highcont-hst}

\begin{figure*}
\includegraphics[scale=1.25]{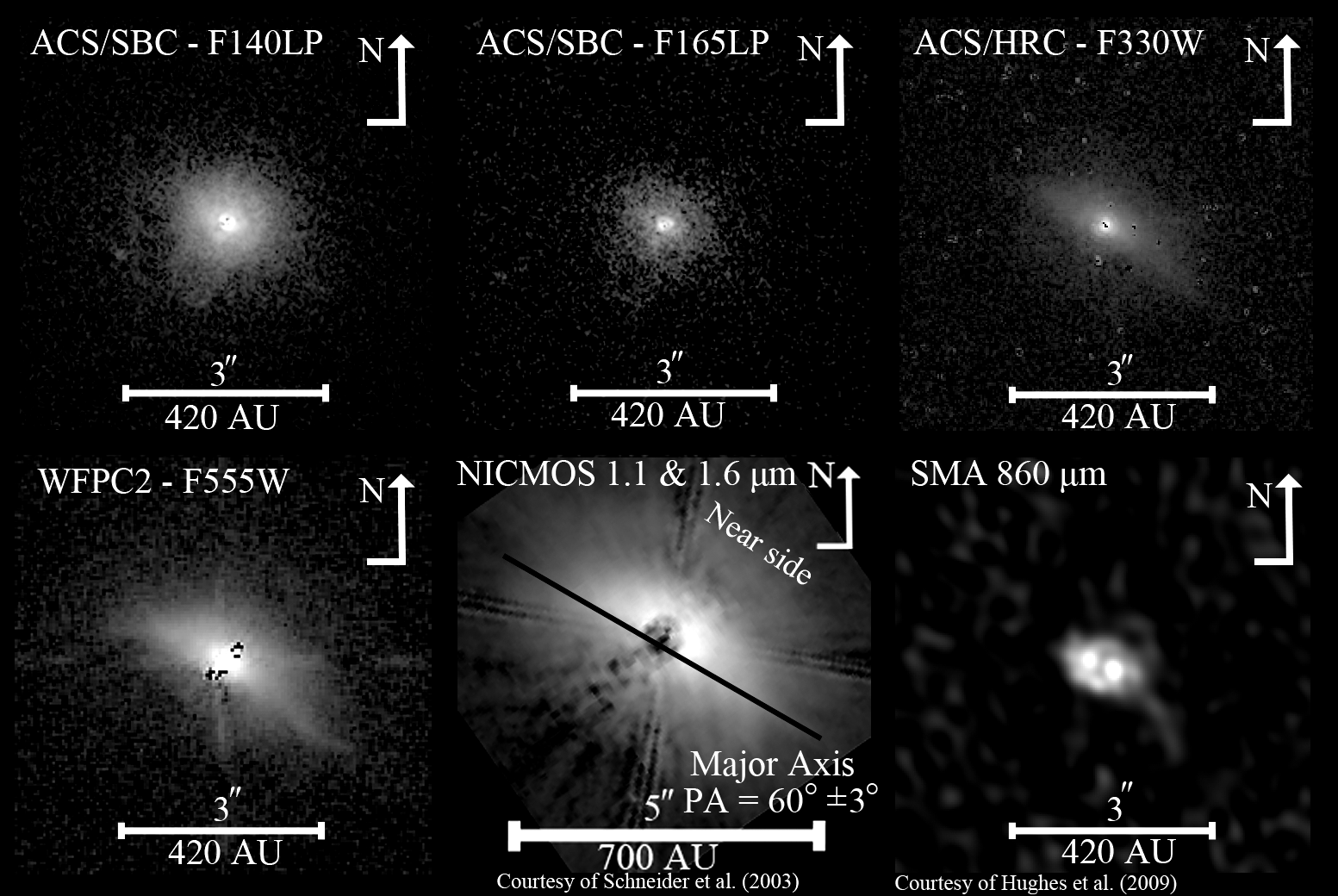}
\caption{\it 
Top Row: \rm\ PSF-subtracted HST imagery observed in the UV with the F140LP
(left), F165LP (center), and F330W (right) filters.  \it Bottom Row: \rm\ PSF
subtracted HST imagery observed in the optical with the F555W (left, from a
single exposure pair, Fig.~\ref{fig-f555w_exposuresubtr_grid}), the F110W \&\ F160W combined 
image from \citet[center,][]{Schneider_03}, and
860~\mum\
from the Submillimeter Array \citep[right,][]{Hughes_09}. All panels
show a $6\arcsec \times 6\arcsec$ FOV and are oriented with north up and east
to the left.  \citet{Hughes_09} noted an inner cavity of size $\sim 20$~AU in
the 860~\mum\ data.}
\label{fig-psfsubtractions}
\end{figure*}

In 1.1 and 1.6~$\mu$m NICMOS imagery, \citet{Schneider_03} detected 
scattered light from the disk of GM~Aur with a major axis along position angle 
PA~=~58.5\degree~$\pm$~2.5\degree, and semi-minor axis in the forward scattering 
direction to be PA~=~328.5\degree~$\pm$~2.5\degree, consistent with sub-millimeter wavelength
observations of the disk \citep{Hughes_09}.  
\citet{Stapelfeldt_95} detected scattered light from the disk 
in WFPC2 bandpasses at 5550~\AA\ and 8140~\AA.  After PSF subtraction, we detect
nebulosity in the optical (3300~\AA) and FUV (1400~\AA~\&~1650~\AA) 
bandpasses (Fig.~\ref{fig-psfsubtractions}).                                            %figure 4 ref
To verify our placement of the major axis for a radial surface brightness
measurement, we constrain
the disk major axis from the images in the F330W and F555W 
bandpasses and determined a PA of 59\degree~$\pm$~3\degree, consistent with the
measurements in \citet{Schneider_03}. The illuminated 
portion of the disk that is detected in each bandpass varies with
exposure depth, so it is likely that we are not 
detecting signal from the full extent of the disk.                                                               
However, the disk is similar in appearance in the optical bandpasses, 
both in and out of bands with emission lines (5550~\AA\ and 3300~\AA\ respectively). 
The consistency of the geometry in the optical bandpasses indicates 
that what we are detecting in these bands is reflection nebulosity.
The measured extent of the disk 
for each bandpass, as well as the aspect ratio, 
can be found in Table~\ref{tab-scatteredlight}, and is                        %Table 2 ref
consistent with a flared disk viewed at an inclination from
pole-on of 56.5\degree~$\pm$~3.5\degree\ \citep{Whitney_92, Stark_06}, 
which we adopt as the disk inclination for this paper.  
Interestingly, the geometry of the disk changes abruptly in the FUV bandpasses.  An
additional component to the nebulosity, aligned along PA = 150\degree\ $\pm$
5\degree, is marginally detected at F165LP and firmly detected in F140LP and will be
discussed in \S~\ref{FUV_structure} in greater detail.  The ACS/SBC and HRC data now extend the
scattered-light detection of the disk of GM~Aur over more than a decade in
wavelength in comparison with the  NICMOS data.

\subsubsection{WFPC2/F555W} 
\label{subsec-wfpc2-f555w}

\begin{figure}[ht]
\centering
\includegraphics[scale=0.5]{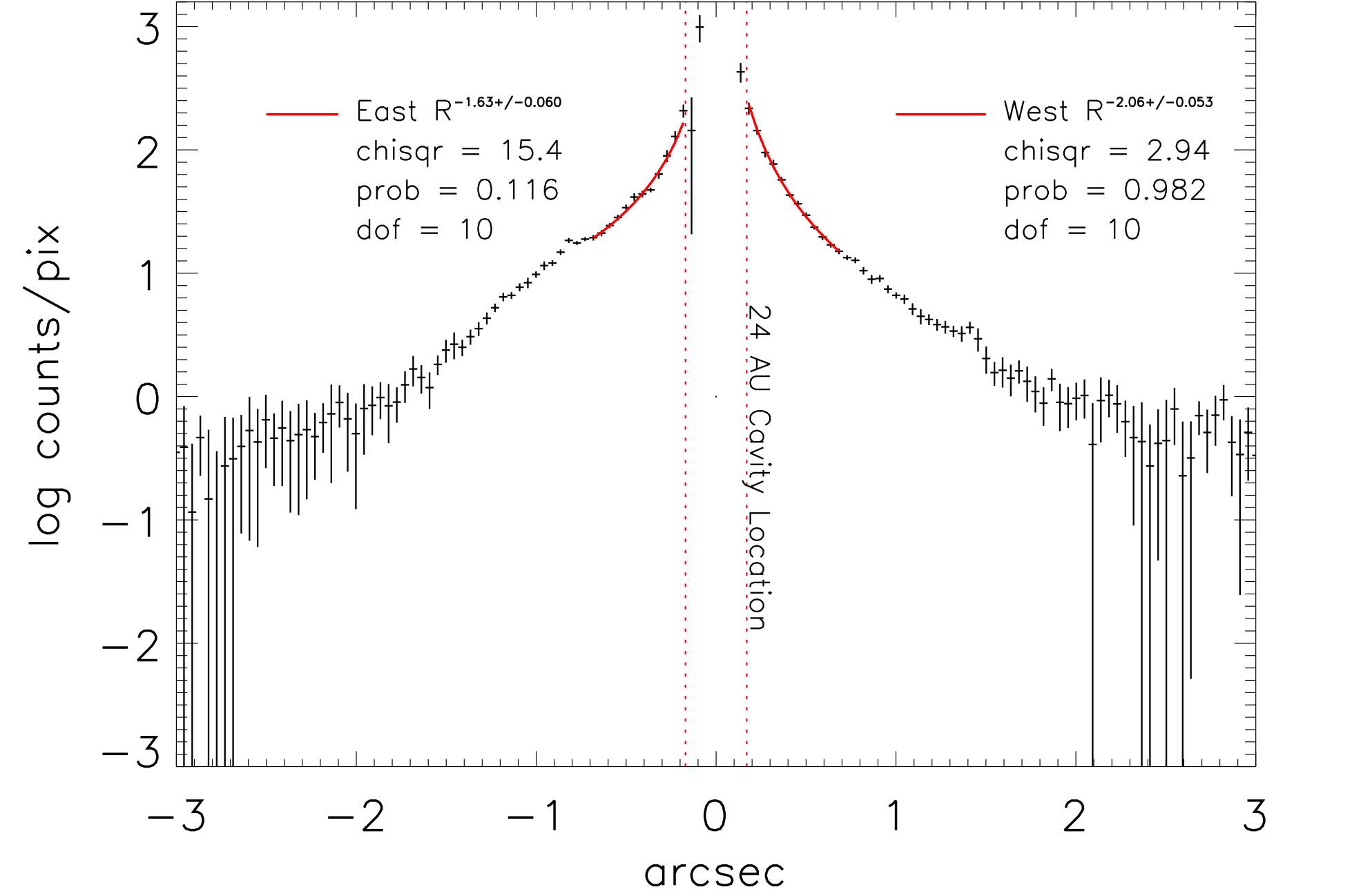}
\caption{The radial surface brightness profile (in log counts pixel$^{-1}$)
as measured from north-east (left side) to south-west (right side) in 
the WFPC2 F555W data along the disk major axis 
PA = 58.8$^\circ$ $\pm$ 3$^\circ$. The resulting counts/pixel 
(surface brightness) {\it vs.} radius relation from 0$\farcs$15 to 0$\farcs$70 was fitted to a
line in log-log space with the IDL {\it linfit} routine, to determine the power law
index $\gamma$ for surface brightness $\Sigma \propto r^{-\gamma}$ (Table~\ref{tab-f555w_surfbrightness}).  The sub-mm cavity wall
falls just within the IWA for the F555W 
profile. There is only one PSF star available, so we cannot show any uncertainty due
to PSF template variations.
\label{fig-555radsurfbrightness}}
\end{figure}

By examination of the F555W PSF-subtracted image, we estimate an 
effective inner working angle (IWA) of 0$\farcs$15-0$\farcs$20
(21-28~AU at d=140 pc).  This is about 1.5 - 2 $\times$ the 80\%
encircled energy radius for the center of the PC chip 
\citep{WFPC2_instrument_handbook}.
To determine the radial surface brightness of the disk in 
the F555W data, we measured the major axis alignment, PA = 58.8$^\circ$ $\pm$ 3$^\circ$, 
then rotated the FITS file image to align the major axis horizontally in the frame with 
the location of the star in the frame center. We then
calculated the median flux value (in counts per pixel) along a three pixel integration
wide strip (0$\farcs$137 or 19~AU at 140~pc) as a function of radius from the flux 
centroid. Using the calculated median $\sigma$ value and dividing by $\sqrt{3}$ 
(for the integration strip width) we estimated the random error.  
We list results separately for 
the east and west sides of the disk. The resulting counts/pixel 
(surface brightness) {\it vs.} radius relation from 0$\farcs$15 to 0$\farcs$70 was fitted to a
line in log-log space with the IDL {\it linfit} routine, to determine the power law
index $\gamma$ for surface brightness  $\Sigma \propto r^{-\gamma}$.  We 
limited the range of our line fit to radii of 0$\farcs$15 to 0$\farcs$70, 
because the fit was affected by a diffuse region of increased flux at
r~$\approx$ 0$\farcs$7 that we cannot securely discount as an artifact.  
As an additional test of the major axis alignment, we varied the FITS 
file rotation angle from the PA by $\pm$ 5\degree\ to constrain 
the variation incurred along varying trial major axis integration strips.  
The west side has a power law index of $\gamma$ = 2.06$\pm$ 0.05, whereas the
east side has $\gamma$ = 1.63$\pm$ 0.06. 
These results are listed in  Table~\ref{tab-f555w_surfbrightness}, and indicate that              % table 3 ref
variations in slope arising from small changes in rotation angle are commensurate
with the random (pixel) errors, at about 0.05-0.1~dex.  In Figure~\ref{fig-555radsurfbrightness}, % fig 5 ref
we plot the radial surface brightness profile, and do not detect a break in the   
surface brightness that would indicate the presence of a cavity wall exterior to 21 AU.
The IWA in the F555W bandpass is not sufficiently small to eliminate 
the possibility of a detectable cavity for the region of the disk interior to 21 AU.   

\subsubsection{HRC/F330W}
\label{subsubsec-hrc-f330w}

The HRC F330W imagery is less heavily exposed than the WFPC2 data,  and we
estimate that it provides an IWA of 0$\farcs$1 (14~AU).  This corresponds to 
$\sim 40-70\%$ of the 80\% encircled energy radius depending on whether
the F330W mode behaves more like the F220W or F435W modes, which are bracketing
modes plotted in \citet{ACS_instrument_handbook}.
Similar to our
F555W analysis, we determined the major axis of the disk to be aligned along PA
= 59.5\degree\ $\pm$ 3\degree\ in the F330W data.   After rotating 
the PSF subtracted FITS file 													 %added "PSF subtracted" JBH 20160415
image to align the major axis horizontally, we calculated the median flux value
(in counts per pixel) along a five pixel integration wide strip (0$\farcs$125 or
17~AU at 140~pc) as a function of radius from the flux  centroid. Using the
calculated median $\sigma$ value and dividing by $\sqrt{5}$  (the integration
strip width) we estimated the random error. Again for comparison, we repeated
the calculation with the FITS file rotation angle changed from the  PA by $\pm$
5\degree .  The radial surface brightness (in counts/pixel) profile index along
the west and east sides is $\gamma$~=~2.17~$\pm$~0.04 and
$\gamma$~=~1.41~$\pm$~0.03 respectively. This is consistent with the east
{\it vs.} west difference in power law index measured in the F555W image. 
Although we optimally nulled the diffraction spikes, the empirically measured
$\sigma$ values are small even compared to any potential Airy ring residuals
(for example at 0$\farcs$4 radius).  This yields an uncertainty in  the power
law fit $\gamma$ value that does not include other sources of error, namely 
uncertainty in PSF subtraction scaling values.  While this does  not alter any
of our results significantly,  the uncertainty in the $\gamma$ values is likely
larger than the formal least squares fit error
calculated by the {\it linfit} IDL routine.

\begin{figure}[ht]
\centering
\includegraphics[scale=0.57]{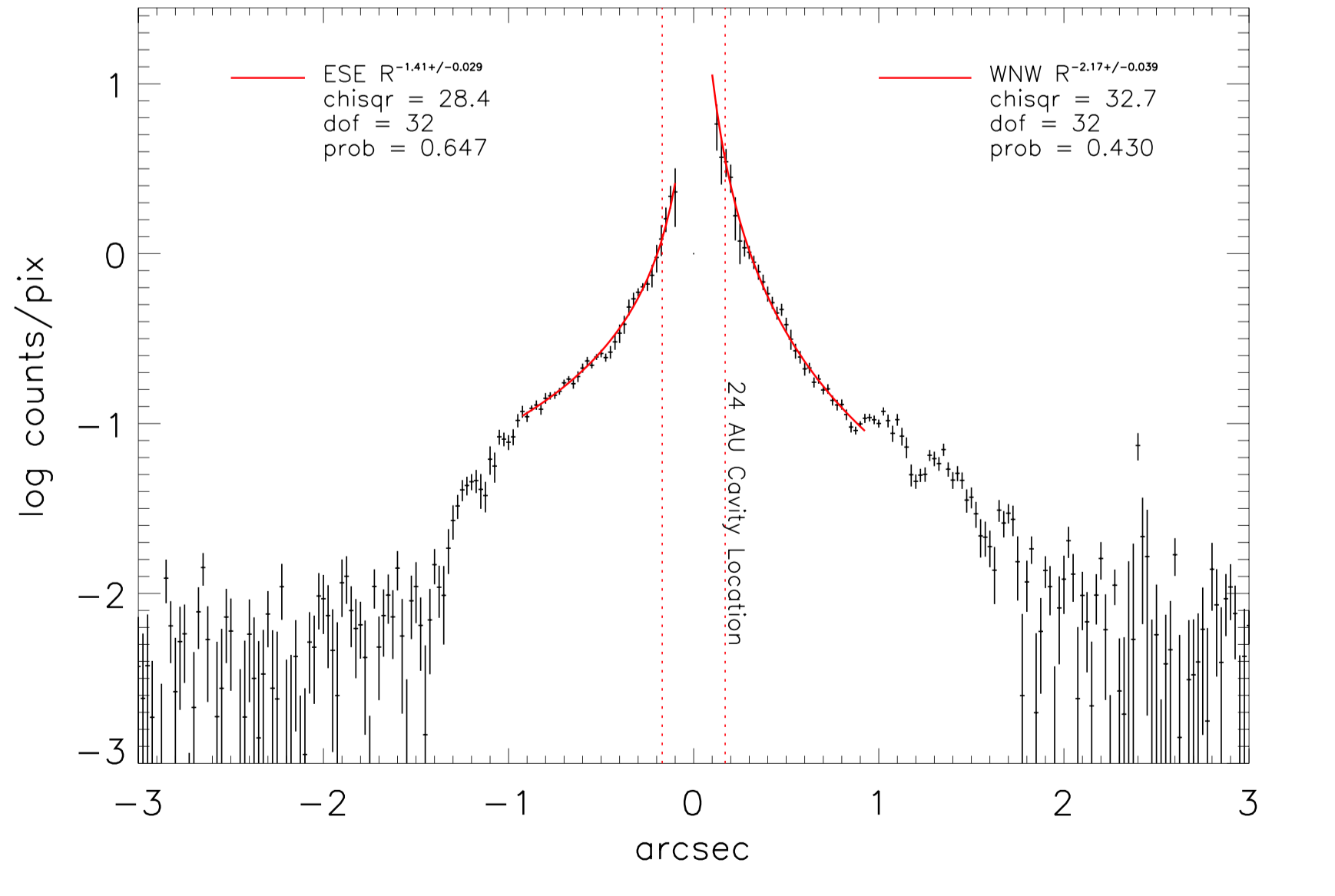}
\caption{
The radial flux profile (in log counts pixel$^{-1}$)
along the disk (north-east to left, south-west to right) major axis 
(PA = 59.5\degree\ $\pm$ 3\degree)
as measured in the ACS/HRC F330W
data.  The best fit to these data are calculated from the inner working
angle (IWA) 0$\farcs$10  to 
1\arcsec\ using the IDL {\it linfit} routine. The 24~AU cavity seen in 
the sub-millimeter \citep{Hughes_09} is not detected in the 
radial brightness profile of the F330W data. 
The main point of this figure is to show the {\it lack} of a break in the
radial surface brightness profile {\it inside} of the purported cavity.
This lack of a break is unlikely to be a consequence of a poor
choice of PSF template.
\label{fig-330radsurfbrightness_data}}
  
\end{figure}

We note no indication of a break in the radial surface brightness profile, and a
smooth power law fit lies  within $\sim 2\sigma$ (measured empirically from
pixel statistics)  of the data points inward to 0$\farcs$1  ($\approx$ 14 AU;
Fig.~\ref{fig-330radsurfbrightness_data}).  This indicates that nebulosity  is  %fig 6
present within the sub-mm cavity of GM~Aur's disk.  The presence of nebulosity
within the sub-millimeter cavity/gap has previously been reported both for disks
around Herbig stars \citep{Muto_12,Grady_13} and T~Tauri stars
\citep{Follette_12} in the NIR. This is the first report of a similar phenomenon
in the optical and FUV.

\subsection{Geometrical Constraints on the Disk in Scattered Light}

Disk geometry can be determined by power-law fits to its radial surface
brightness (SB) profile \citep{Grady_07,Wisniewski_08}.   Geometrically flat or
minimally flared disks with an outer wall structure, where the walls do not
shadow the outer disks, result in power law fits of $\approx r^{-3}$ to their
radial SB profile (e.g., HD~100546, \citealt{Grady_01}; HD~169142,
\citealt{Grady_07}; HD~97048, \citealt{Doering_07}; HD~163296,
\citealt{Wisniewski_08}).  In contrast, the disk  of GM~Aur has the morphology
of a flared disk seen in scattered light, i.e. brighter along the forward
scattering (north side) semi-minor axis, and  bilaterally symmetric about the
minor axis in surface brightness distribution \citep{Whitney_92, Stark_06}.  The
power-law fit to its radial surface brightness profile of $\approx r^{-2}$  is
further evidence that the disk of GM~Aur is flared.    

The surface brightness profile of the disk from the ACS/HRC F330W data also
lacks any evidence of the 24~AU cavity that was observed by \citet{Hughes_09} at
sub-mm/mm wavelengths.  The non-detection of a change in the power-law fit to
the surface brightness profile at or near 24~AU,
and the fact that {\it we do not see diffraction spikes} 
indicate that the disk is
optically thick inside this radius at  these wavelengths. 
We conclude that in the far-UV we see
scattering from particles on the order of the observed
wavelengths \citep{Whitney_92}, where the morphology displays
the characteristic pattern of a flared disk.    
We therefore infer that we observe light scattered by sub-micron
sized grains within the 24~AU sub-mm cavity.
This conclusion was also reached by \citet{Calvet_05}.

\subsection{Grain Properties}
\label{subsec-grainproperties}

Monochromatic imaging is insufficient for characterizing a particle size
distribution because the relationship between the grain opacity/albedo and
particle size is degenerate \citep{Watson_07}.  However, multi-wavelength
imaging can break this degeneracy if done over a sufficiently wide wavelength
range.  Our data extend the wavelength coverage for GM~Aur by an order of
magnitude.  \citet{Schneider_03} modeled the disk of GM~Aur exterior to
0$\farcs$3  (thus exterior to a 24~AU disk) and were able to accurately
reproduce the size, overall brightness level, large-scale surface brightness
distribution, etc. using a grain model with a maximum radius of 1~mm, an
exponential cutoff scale length of 50~$\mu$m \citep[Model 1 in][]{Wood_02}, and
a minimum grain size of 0.05~$\mu$m (Wolff \&\ Wood priv. comm.). 
\citet{Schneider_03} measured the disk scattering fraction, the  ratio of light
scattered by the disk to starlight (\fdiskstar ).  We also measure \fdiskstar ,
(1) because it is an extinction-free parameter, (2) because it is a  measurement
of the disk albedo, (3) to determine how \fdiskstar\ varies as a
function of wavelength, and (4) to compare the result with grain opacity models 
and constrain possible grain compositions.  We note that as \fdiskstar\ applies
to the disk surface, at distances of $\sim50-1000$~AU, so any short-term
($<0.5$~day) variability from flares should not affect it significantly.

\begin{figure}
\includegraphics[scale=1.0]{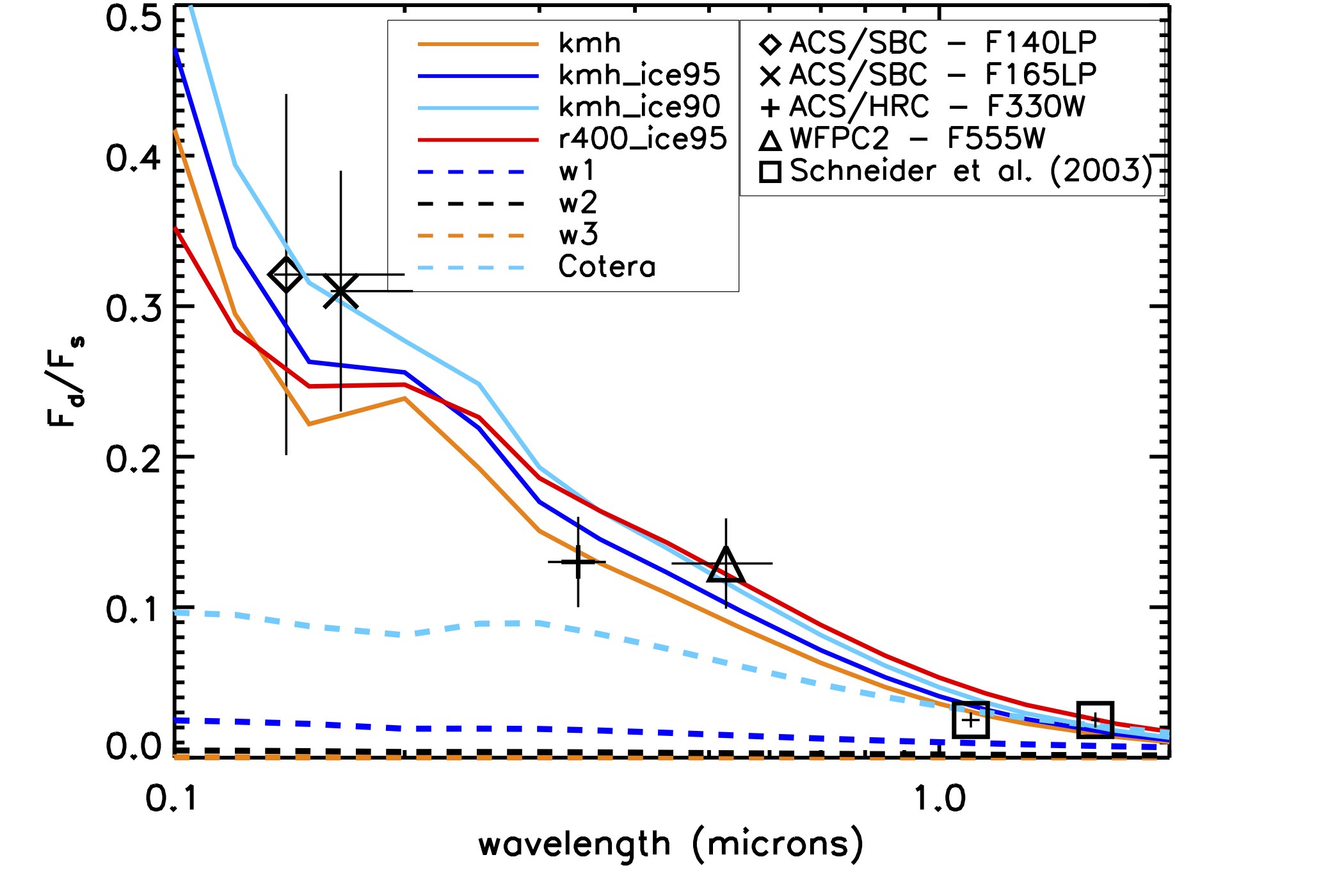}
\caption{The disk scattering fraction at the disk surface
$F_{disk}/F_{star}$ as a function of wavelength has been measured and
compared with grain opacity models that have  varying grain size distributions. 
The  w1, w2, w3, and Cotera models are model 1, 2, 3, and Cotera from
\citet{Wood_02} respectively, and  they have size distributions dominated by 
large (micron-sized) grains or larger and result in gray scattering (flat
curve).  The r400\_ice95 model is a standard one offered in the 2014
version of {\sc Hochunk3d} \citep{Whitney_13}, adopted
from \citet{Whittet_01} and modelling total-to-selective
extinction of $R_V=4$, corresponding to dense regions of 
star-forming clouds with optical depth $\tau_{3.0}$
for the 3~$\mu$m ice absorption feature.
The r400 model and the kmh models have size distributions similar to
that of the ISM,  with variations in composition, and their scattering
efficiency increases rapidly at shorter wavelengths. See
\S~\ref{subsec-grainproperties}   for additional details.}
\label{fig-fdiskstar}
\end{figure}

If the dust grain model from \citet{Wood_02a} used 
in \citet{Schneider_03} is correct, the disk 
dust opacity would exhibit a shallower wavelength-dependent
opacity than the canonical interstellar medium extinction curve
\citep{Cardelli_89}.  To examine the wavelength-dependence of the resultant
scattered light,
we determined \fdiskstar\ for the
F140LP, F165LP, F330W and F555W images as follows.  For the disk flux,
we measured aperture photometry from 0$\farcs$3 out to a radius of 3$\farcs$4, with a 
sky annulus of 4-6\arcsec , in the PSF-subtracted image.
Results were not sensitive to the choice of the sky annulus.
For the stellar flux comparison, we measured the flux in the entire 
region out to 3$\farcs$4 in both the scaled PSF template data and in the
GM~Aur data without PSF subtraction.  The stellar flux strongly dominates
over the disk flux.
We then determined \fdiskstar\ uncertainties 
for the F330W and F555W bands, where there are diffraction spikes, based on
the PSF subtraction scale factor uncertainty (the 
dominant uncertainty in the flux).
This was set by requiring that the residual
diffraction spikes be within $1\sigma$ (rms) of the local mean (\S~\ref{subsubsec-data-f330w}). 
For the F140LP and F165LP bands (discussed in detail
in \S~\ref{FUV_structure} and \S~\ref{subsec-possmolecularoutflow}), we constrained
the uncertainty in \fdiskstar\ 
using the lower and upper bounds of the scale factor used for the PSF template. There are
no diffraction spikes in these data (see \S~\ref{subsubsec-data-f140lp} for details).
We supplement our values with measurements for the F110W and F160W bands
from  \citet{Schneider_03} of 0.025 $\pm$ 0.005 (20\%)\footnote{\citet{Schneider_03} also 
give a value of $\approx$ 4\%.  After private communication, we learned that 
this is an error and should have read $\approx$ 2.5\%. }.  
Values for \fdiskstar\ are listed in
Table~\ref{tab-scatteredlight}, and plotted in                                                                                                                  %table 2 ref
Fig.~\ref{fig-fdiskstar}.  This figure also                                                      %fig 7 ref
shows dust grain opacity curves that represent a
range of grain size distributions and compositions. 

The grain opacity profiles in
Fig.~\ref{fig-fdiskstar}   have been scaled to fit our measurements and those of                                                %fig 7 ref
\citet{Schneider_03}.   The grain opacity models kmh, kmh\_ice95, and kmh\_ice90
originate from \citet{Kim_94}, and are representative of ISM-like grains. The
\_ice\#\# extension indicates a water ice coating on the grain surface. The
number indicates the ratio of grain material to water ice, where ice90 is 90\%\ grain
10\%\ water ice coating on the grain surface, ice95 is 5\%\ water ice, etc.  
Models 1, 2, 3, and Cotera  in \citet{Wood_02} correspond to w1, w2, w3, and Cotera respectively in
this paper, and are grain models dominated by  large dust grains.  These
large grain models are described in detail in \citet{Wood_02} and \citet{Cotera_01}.
All models in \citet{Wood_02} are composed of amorphous carbon-silicon, and they are dominated by
micron-sized grains or larger (min = 0.05~$\mu$m, max = 1~mm).  We rule out the w1, w2, w3, and Cotera
models as  candidates for the
surface grain composition of GM~Aur, at approximately the 4~$\sigma$ level 
for Cotera grain model, and at the 7~$\sigma$ 
level for the others, based on our scattered light 
measurements of the disk in the F140LP, F165LP, F330W and F555W data 
(Fig.~\ref{fig-fdiskstar}).  To compare the disk scattering fraction with the 
grain opacity of the models, we scaled the grain models to fit the data points 
given in \citet{Schneider_03} and the F330W and F555W data, because the FUV data have larger errors.  
The kmh models and the r400 model all lie close to the values for \fdiskstar\ we have measured, but
the uncertainty in our values is large enough that we cannot decide conclusively between these
models.  However, the only grain model with an opacity curve that fell inside the error bars of all of
our \fdiskstar\ measurements was the kmh\_ice95 model.
The wavelength dependance of the disk-scattering fraction we find
indicates that 
the disk surface is populated with grains that have a size distribution similar to that
of the ISM, both in
the outer disk and within the region of the sub-millimeter cavity.  It is
possible that the physical grain population at the surface of the disk differs in
composition from the grain models we have presented.  However, if so, we can
exclude pure frosts of H$_2$O, CO$_2$, NH$_3$ and SO$_2$.  The reflectance measured
for each of these pure ices is flat in the optical and slightly rising to its peak at
$\approx$ 0.2~$\mu$m (except SO$_2$ which remains flat) before each drops rapidly to 
its minimum reflectance at $\approx$ 
0.16~$\mu$m \citep{Hapke_81}. These same disk models were then used to 
construct a synthetic image, discussed in \S~\ref{subsec-possmolecularoutflow}.

\subsection{Structure Unique to the FUV \label{FUV_structure}}

Initial inspection of the PSF-subtracted ACS/SBC data unexpectedly revealed an
extended structure that is not detected in any of the other bandpasses.  
The disk seen at these shorter wavelengths appears
cylindrical in shape and limb-brightened (Fig.~\ref{fig-psfsubtractions};                                                                       % fig 4 ref
top left \& top middle panels).  
The structure is detected strongly in
the F140LP bandpass, containing roughly $\sim 10$\% of the flux in the star+disk.
It is also seen
marginally in the F165LP bandpass, and can be
traced out 1$\farcs$1 $\pm$ 0$\farcs$2 along PA $150\pm 5^\circ$  away from
the disk midplane (along the southeastern semi-minor axis).   
De-projecting our measurements to
determine a physical size of the structure, we find that it extends to a
height $190\pm35$~AU above the disk mid-plane.  The wall thickness of the
feature, or equivalently the area where we detect limb brightening, is 0$\farcs$17$\pm$0$\farcs$05
or $24\pm 7$~AU.  The radial brightness of the feature, measured along the area
where it is limb-brightened, decreases as r$^{-0.7 \pm
0.1}$.  We measure the outer radius of the feature to be 0$\farcs$46
$\pm$ 0$\farcs$05 ($64\pm 7$~AU).  From this we calculate that the inner radius
is $40\pm 10$~AU, which corresponds to the inner region of the outer disk just
beyond the radius of the sub-mm cavity wall.  The abrupt appearance of this
structure in the ACS/SBC data is consistent with the source being 
seen in emission. 

As a transitional disk, GM~Aur is beyond the evolutionary stage where an outflow
cavity with a narrow opening angle would be expected \citep{Seale_08}, and the
cylindrical geometry of the structure is inconsistent with the parabolic or ``V''
shaped geometry that is characteristic of outflow cavities in younger systems.   This
shape is also inconsistent with the typical appearance of an atomic or ionic jet,
i.e. a collimated signal aligned along the minor axis. 
The main optical jet diagnostics are high-velocity emission in the lines of $[$\ion{O}{1}$]$~6300~\AA ,
$[$\ion{N}{2}$]$~6580~\AA\ and $[$\ion{S}{2}$]$~6731, 6716~\AA\ and H$\alpha$. In GM~Aur,
there is no evidence for a high-velocity component (indicative of a collimated
jet) in these lines \citep{Hartigan_95}.  We see no such collimated signal
\citep[also see][]{Cox_07}.

Another argument against an ionic jet would be that the forbidden emission
line $[$\ion{O}{3}$]$, which sometimes may be found in jets (albeit
requiring high velocity shocks of $v>100$~km~s$^{-1}$),
would be detected in the F555W bandpass \citep{WFPC2_instrument_handbook} as a
collimated source aligned along the system minor axis.  Again,
no such collimated signal is                                                            
detected in the WFPC2 data, and the structure detected in the SBC
data does not exhibit the typical jet morphology.  
The fact that the structure is not
detected in a bandpass dominated by reflection nebulosity indicates that the source
is not predominantly dust. 
It is unlikely that very small grains (transiently heated)
could produce the diffuse UV emission, because the flux from the star is 
insufficient.  In addition, the small grain thermal emission would have the
morphology of the disk -- which is not the case for UV emission around GM~Aur.
The source of the emission can be investigated more
thoroughly with long-slit spectroscopy; 
we discuss the possibility
that the emission may arise from H$_2$ in \S~\ref{subsec-possmolecularoutflow}.

\subsection{SED and Model Image \label{SED_model}}

\begin{figure}
\includegraphics[scale=1.0]{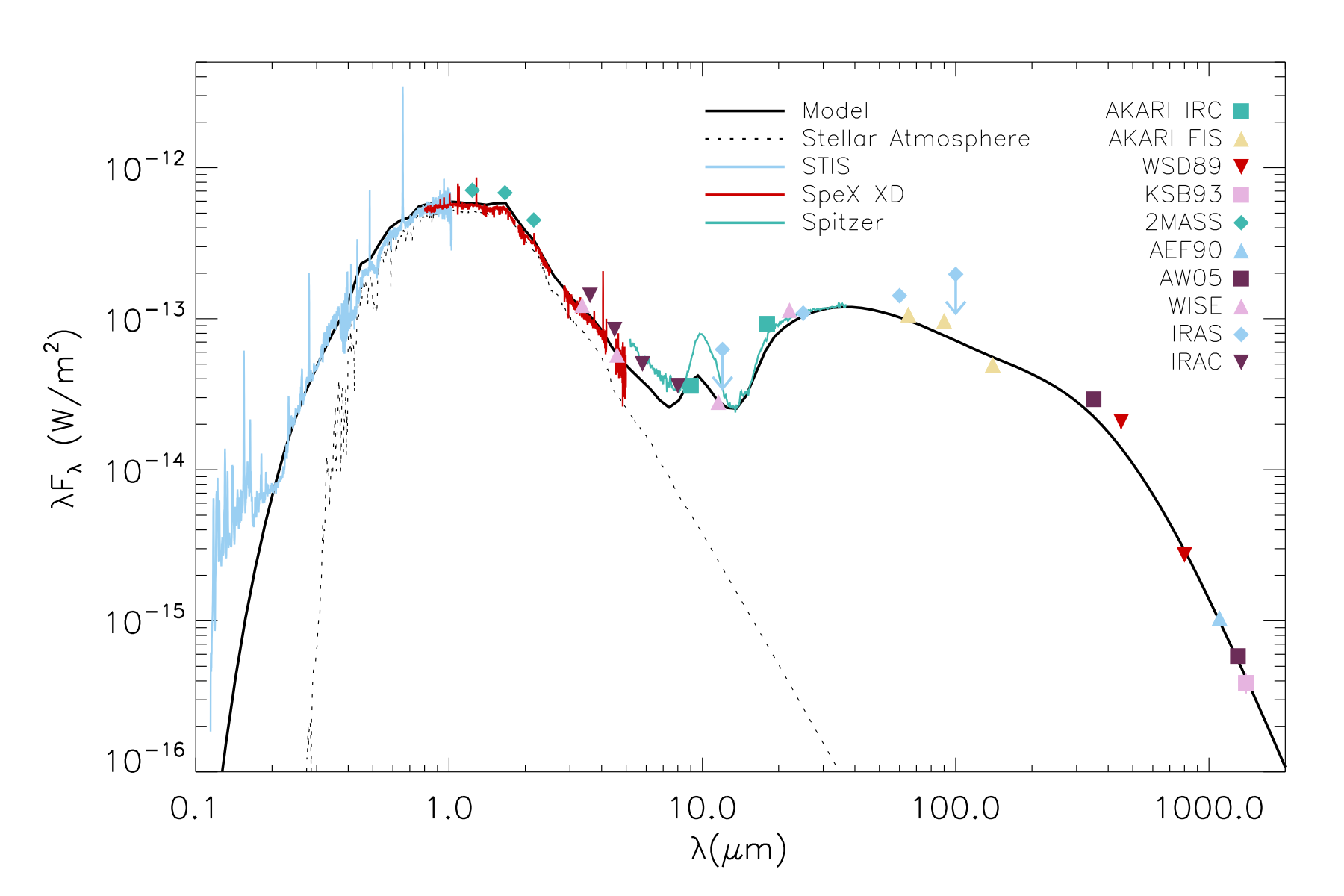}
\caption{The SED of GM~Aur using data  discussed in 
\S~\ref{subsec-otherdata}, and best-fit MCRT model (\it{solid black line}; see \S~\ref{SED_model}\rm). The X-ray flux is
$6.82\times 10^{-16}$~W~m$^{-2}$ at $\lambda\approx 0.0021$~$\mu$m \citep{Gudel_10},
and would be consistent with an extrapolation of the UV flux; we do not plot it as it
would crowd the other data points into one quadrant of the panel.  Paper references
are: WSD89 = \citet{Weintraub_89}; KSB93 = \citet{Koerner_93}; 
AEF90 = \citet{Adams_90}; AW05 = \citet{Andrews_05}.  The SPEX and STIS data 
are from 2011.}
\label{fig-sed}
\end{figure}

We modeled the spectral energy distribution and images of GM~Aurigae using the
2013 version of the Whitney Monte Carlo radiative transfer (MCRT) code
\citep{Whitney_13}. The SED was fitted to the data mentioned in 
\S~\ref{subsec-otherdata}
(Fig.~\ref{fig-sed}).                                            %fig 8
Fitting a model SED to an observed SED is not conclusive on its own due to the
degenerate nature of the SED inherent from the vast parameter space
\citep{Chiang_01,Robitaille_07}.  Comparison of model data to the corresponding
observational imagery and our F$_{disk}$/F$_{star}$ measurements can reduce 
the
number of free parameters.  The stellar parameters were largely determined by the
value of $A_V=0.1$ we chose based on the extinction stated in \citet{France_14}; 
a  larger value would have been less consistent with the FUV flux.  
\citet{Hughes_09} give $T=4730$~K, $R=1.5R_\odot$, $d=140$~pc and $A_V=1.2$, 
\citet{Hueso_05} use $T=4060$~K, $R=1.83R_\odot$ and $d=140$~pc and
\citet{Pott_10} employ  $T=4730$~K, $R=1.5R_\odot$ and $A_V=1.2$. Our best fit
model to the SED shape indicated parameters of $T=4000$~K, $M=1.2M_\odot$,
$R=1.5R_\odot$ and $A_V=0.1$. The presence of FUV flux supports the lower value of
$A_V$ we choose, which requires a cooler star to give the correct stellar
contribution to the SED.  We employ a disk cavity size out to 24~AU 
\citep{Calvet_05, Hughes_09, Grafe_11, Andrews_11}.   The only free parameter left
for the star is the distance, which scales the model spectrum. This constrains
$d=148$~pc for our models, which is within the uncertainty of the
distance \citep{Bertout_06}.  We find evidence for variability in the near-IR and UV portions of the
SED (Fig.~\ref{fig-sed}). The high accretion state corresponds to the 2MASS data
and Spitzer era observations (IRAC  from 2004, IRS from 2005). The variability is
described in detail in \citet{Ingleby_15}.

\begin{figure}[ht]
\centering
\includegraphics[scale=0.05]{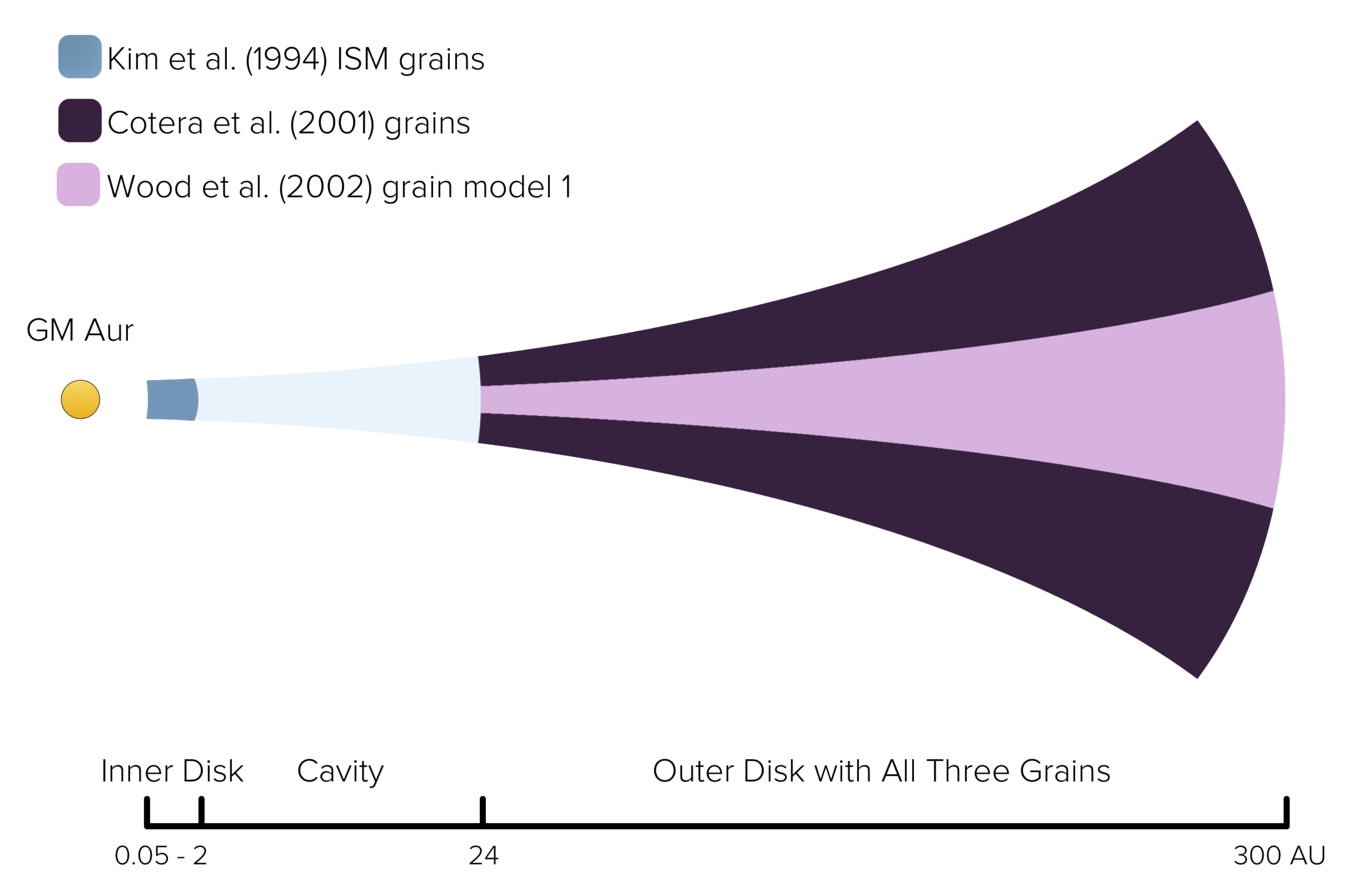}
\caption{The MCRT code from \citet{Whitney_13} was used to model the disk of 
GM Aur, and this schematic illustrates the basic parameters involved.  There is an inner disk 
that extends from 0.5 AU to 2 AU, then a gap devoid of material extends out to 24 AU, 
followed by a flared outer disk with two dust components.  The disk midplane is dominated by 
large grains used to model the edge-on disk of HH30 \citep[Model 1;][]{Wood_02}. Above the midplane, the outer disk has two components. The grains described by \citet{Cotera_01} have a size distribution smaller than the grains at the midplane, but larger than ISM-like grains. This 
region also contains a population of ISM-like grains {\it (not shown via shading)}
and is described by \citet{Kim_94}. \label{gmaur_schematic}}

\end{figure}

The disk itself was modeled to match the observed SED 
with three components (Fig.~\ref{gmaur_schematic}). The first is an %fig 9 ref
inner disk composed of the grains of \citet{Kim_94}, 
primarily fixed by the 10~$\mu$m strong silicate feature we see in the data,
which extends from the sublimation radius out to 2~AU. 
This inner disk is composed of kmh grains \citep{Kim_94} 
and comprises $4\times 10^{-7}$
of the overall disk mass, which is 0.092~$M_\odot$ {\it in total}. 
The r = 2~AU radius for the inner disk is motivated 
by the model used by \citet{Calvet_05}, whose inner disk region was less than 5~AU. 
This size is consistent with the 10~$\mu$m emission
\citep{Hughes_09}, and
is otherwise optically thin in the IR as was
the model disk for \citet{Calvet_05}.  It is not necessarily optically thin 
in the UV and at blue optical wavelengths.  In that case, in the more 
optically thick UV we see the outer 
layers of the disk, while at longer, more optically thin wavelengths we see 
deeper into the disk \citep[e.g. see their Fig.~6 caption]{Pinte_08}.

The second
component was a disk settled near the mid-plane composed of the grain
prescription of \citet{Wood_02}, to match the SED at mm wavelengths. 
The disk contained 0.07~$M_\odot$ {bf (i.e. $\sim 80$\%)}
of the total disk mass and extended from
24~AU to an outer disk radius of 300~AU. This grain type and mass
fraction were chosen in order to fit the mm wavelength data
effectively. Lastly, a second outer disk extending vertically off the
mid-plane was composed of the grains of \citet{Cotera_01}, also
extending from 24~AU to 300~AU. 
The Cotera grains were adopted because they best fit the SED 
complementary to the mm region produced 
by the outer disk regions. We know something kmh-like must exist in 
the inner disk region given the very strong silicate feature, and for 
simplicity's sake, a disk comprised of kmh grains throughout was attempted. 
However, this just did not produce a good fit within the other constraints 
of the model. In the end, a larger grain settled disk comprised of Wood et al. 
model 1 grains with a Cotera jacket and some kmh grains throughout produced the best fit. 
Other grain composition/distribution models may also work. The icy kmh
grains are the best fit to the observed $F_{disk}/F_{star}$. The amount of
allowed parameter space for grain types and distribution characteristics
may be large, and we defer more detailed modelling to a subsequent paper.

Two hotspots were also added on 
the star: one facing towards and another away from the observer. The hotspots along with an 
accretion rate of $\dot{M}$ = 4 $\times$ 10$^{-9}$ M$_\odot$ yr$^{-1}$ served as a source
of FUV emission needed to fit the STIS data set portion of the SED. The code
was run with $2\times 10^9$ photons for the image models, which was the maximum
permissible in the version of the code available.  

The Whitney code comes with predefined filters that mimic the throughput of filters 
commonly used in observations.  Our novel approach is the first to require FUV filters, 
which are not included with the code.  The code does, however, allow the creation and
implementation of user-defined filters. We thus added filters using the throughput
curves for the F140LP \& F165LP filters in the ACS Instrument Handbook
\citep{ACS_instrument_handbook}.
We convolved the model images with a Gaussian filter ({\sc gfilter}), 
from the IDL IUELIB astronomical 
library.\footnote{www.astro.washington.edu/docs/idl/cgi-bin/getpro/library38.html?GFILTER}
The FHWM of the PSF star images for F140LP, F165LP and F330W were measured
with the IRAF {\sc imexam} routine\footnote{IRAF is distributed by the
National Optical Astronomy Observatories, which are operated by the 
Association of Universities for Research in Astronomy, Inc., under
cooperative agreement with the National Science Foundation}, yielding
Gaussian approximations of 0$\farcs$10, 0$\farcs$09 and 0$\farcs$06, respectively.  We used
half those values as the standard deviation for the {\sc gfilter}
input, using enough points to go out $\sim 4$ standard deviations
when filtering the model images.  We then used the {\sc artdata} routine to add noise
to the F330W model, and background pixels from the SBC data for 
the F140LP and F165LP model images.  Finally, we subtracted the
appropriate PSF templates from the images.

\begin{figure}
\includegraphics[scale=0.50]{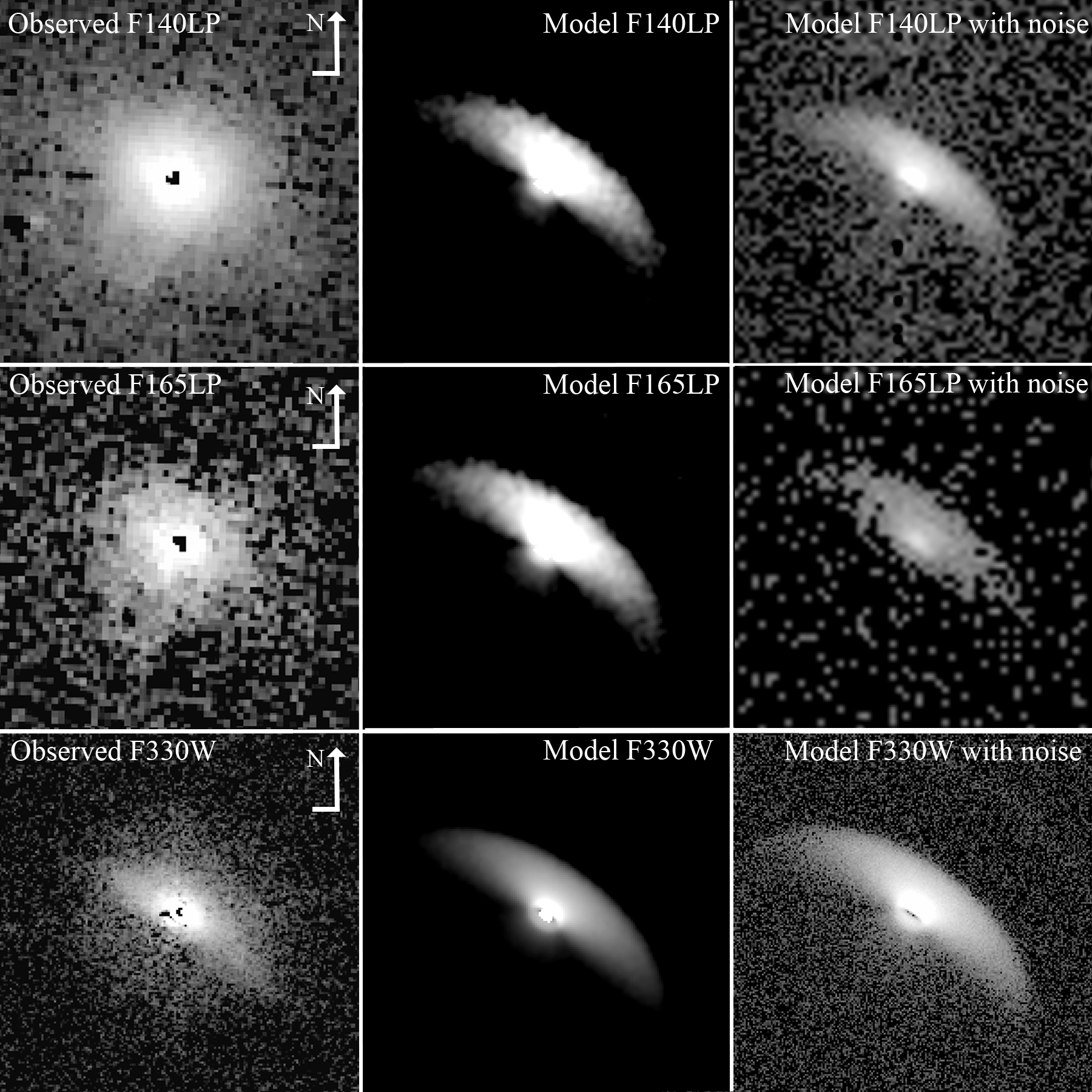}
\caption{Comparison of PSF-subtracted ACS images for GM~Aur 
(left panels) and Whitney code models rotated and scaled to match
the images (right
panels) for F140LP, F165LP and F330W (top to bottom).  Field sizes are 600~AU
for the models, and 4$\farcs$0 for the images (600~AU at 148~pc).  North is up and
east to the left in all panels. The models and the images generated from them
are consistent with the GM~Aur SED (see text for details), with
the possible exception of accounting for all of 
the UV excess (involving F140LP and F165LP).  The emission to the SE of the disk
in those two filters is apparent in the data, but not in the models.}
\label{fig-whitney_data_compare}
\end{figure}

\begin{figure}[ht]
\centering
\includegraphics[scale=0.5]{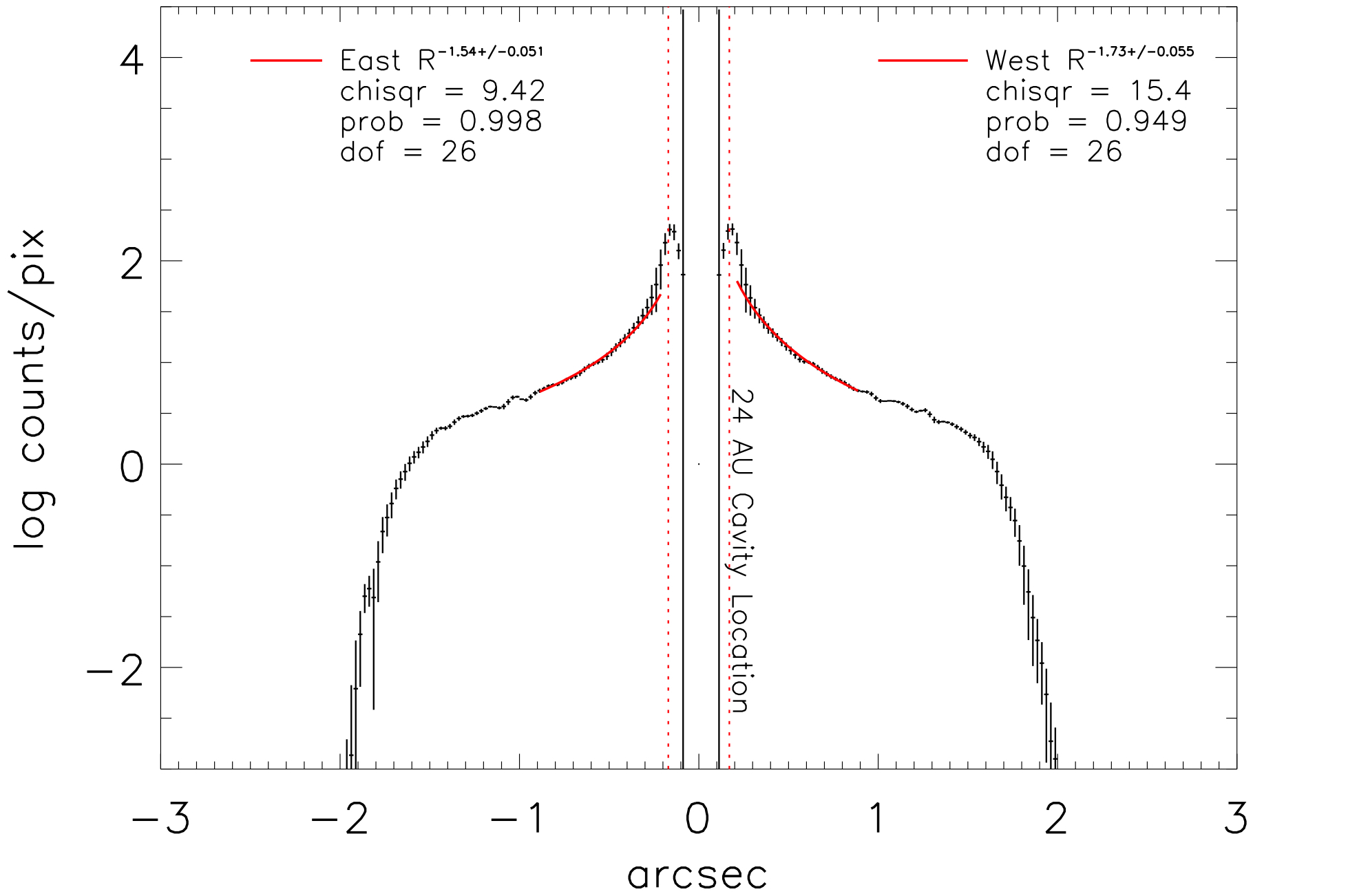}
\caption{ The same radial brightness profile calculation 
as in Fig.~\ref{fig-330radsurfbrightness_data} on the model F330W image.  The 24~AU cavity seen in 
the sub-millimeter \citep{Hughes_09}, which was not detected in the F330W data, is detected it in 
our model image data.  Consequently, the best fit to the model image
required a starting point exterior to the cavity location; the fit from 0$\farcs$2 to 0$\farcs$9 is 
shown here. \label{fig-330radsurfbrightness_model}}

\end{figure}

For the F330W model, the image appears to match the data reasonably well outside of a
radius  of $r \sim 0.2$~arcsec
(Fig.~\ref{fig-whitney_data_compare}).         						%fig 10 ref 
Interior to that point, the gap and inner rim of the disk are visible in the
F330W model image, whereas we find no indication for any gap or rim in the  actual image
data, nor from the radial surface brightness profile from the data
(\S~\ref{subsubsec-hrc-f330w};
Fig.~\ref{fig-330radsurfbrightness_model}).                                     %fig 11
The F165LP model image bears less resemblance to the image data than for the F330W
case, and the data appear much more circular than the model due to emission to the SE.
For the F140LP model image, the difference is even more pronounced,
with the cylindrical projection form to the SE showing more
clearly in the observed data, plus perhaps a smaller amount of 
diffuse emission to the NW.  Although the disk morphology is
well-fitted at wavelengths $\lambda >3300$~\AA , 
we cannot model the observed far-UV emission
well using only scattered light.  There is an extra emission component along the
minor axis which could be consistent with an outflow, which is not included
in the model.

\section{Discussion}

The disk of GM~Aur has now been detected in scattered light from 0.1450~$\mu$m to 
1.6~$\mu$m, more than a decade in wavelength coverage.  This large wavelength lever
allows us to explore the surface dust grain opacity as a function of wavelength,
in a manner similar to \citet{Pinte_08} but with the addition of 
FUV wavelength coverage. In 
contrast to \citet{Schneider_03} who inferred a dust grain composition similar to Model
1 in \citet{Wood_02}, which is composed of large grains (up to 1~mm),  we find that the
wavelength dependence of F$_{disk}$/F$_{star}$ is consistent with ISM-like grains.  This
indicates that little grain growth has occurred in the upper  layers of the disk.
ISM-like grains are also consistent with the radial surface brightness profile of the
disk at 3300 \AA .  The overall color of the dust disk surface is blue 
($m_{F330W}-m_{F555W}~\approx~-0.2$), and the presence of ISM-like grains is
undoubtedly  a factor in disk detection in direct imaging using WFPC2
\citep{Stapelfeldt_95}.

\subsection{Cavity Non-Detection at Short Wavelengths}

The cavity detected in mm, sub-mm, and mid-IR wavelengths 
\citep{Calvet_05,Hughes_09,Grafe_11, Andrews_11} is large enough in radius to be 
detected in the FUV and optical F330W bandpasses.   The consistency in detections
of the cavity at mid-IR and longer wavelengths makes it safe to assume there is 
a cavity in the large grain dust.  However, we report a non-detection 
of the cavity exterior to 15 AU at optical and FUV wavelengths.  Our analysis provides 
no evidence for a break in the radial surface brightness profile that would indicate 
a depletion in the surface density of the grains or signal the presence of a change in 
disk composition.  This can only be the case if the small grains at the surface of the 
disk persist inside the radius of the sub-millimeter
cavity wall. Our non-detection of the cavity indicates that 
there is a mechanism in place actively filtering 
the material within the cavity, allowing only sub-$\mu$m grains to migrate inward 
beyond the cavity wall. (This is not precluded by the SED fit, and does
not conflict with a statement by \citet{Calvet_05} on the outer boundary of the
optically thin region in the IR and/or redward.  
The inner region is optically thick in the far-UV.)
This conclusion is supported by the analysis of ro-vibrational 
CO lines in \citet{Salyk_11}. Their models suggest that an inner disk of CO gas 
extends out to a radius of 0.2 AU of GM~Aur, and that it is likely being replenished 
via gas migration through the cavity from the outer disk.  
This is consistent with Monte Carlo Radiative Transfer and 
hydrodynamical model predictions in \citet{Paardekooper_06a, Rice_06,Zhu_12,Dong_12a} 
\&\ \citet{de_Juan_Ovelar_13}, 
which suggest giant planet formation within the disk cavity as a likely
culprit in this scenario. There are other mechanisms which can also preferentially
decrease the density of micron-sized grains. Grain growth is one
possibility, although it could
be expected to vary smoothly with radius  \citep{Cieza_12}. 
\citet{Dullemond_05} concluded that if grain growth were responsible for cavity
development, small grains must be replenished, possibly by aggregate fragmentation via
high-speed collisions.
\citet{Owen_11} suggested that X-ray photo-evaporation could explain a large
fraction ($\gsim 50$\%) of transitional disks.  However, \citet{Alexander_06b}
suggested that GM~Aur in particular had too high an accretion rate to have its
cavity produced by photo-evaporation, and that it was rather caused by
another mechanism such as planet formation or grain growth/coagulation.
The detection of small dust grains within the millimetric cavity seems to rule out
photoevaporation and dust coagulation processes as the main origin for the
mm feature, as in both cases small dust particles would not be expected inside the hole.
Our result here confirms the predictions of \citet{Calvet_05} and \citet{Espaillat_10}, using
an independent data set.

\subsection{Limits on Giant Planets in the Disk of GM~Aur}

Grain filtration occurs when larger grains are restricted to the outer disk and small
grains, which are more tightly coupled to the gas, can freely penetrate the cavity
\citep{Rice_06}.  Grain filtration is also a predicted consequence of the 
presence of one or more giant planets.   
Based on the models presented in \citet{Zhu_12}, a 3~$M_J$
object would clear a gap in the 30~$\mu$m grains on a timescale of 10$^5$ yrs, and
create a noticeable depletion ($\approx$ 3 orders of magnitude) in the gas density
within the cavity. The models presented in \citet{de_Juan_Ovelar_13} produce
similar results to those presented in \citet{Zhu_12}.  
They find that a 1 $M_J$ planet does deplete 1~$\mu$m sized grains
at the surface of the disk that could be detected in optical ($0.65~\mu$m) 
observations \citep[Fig.~3]{de_Juan_Ovelar_13}, given high enough spatial resolution.  
They find that a 9~M$_J$ planet depletes the 1~$\mu$m grains at the disk surface
by a factor of 1000 \citep[Fig.~7]{de_Juan_Ovelar_13} at the planet location;
a 15 $M_J$ planet eliminates dust grains of all sizes entirely from the disk 
surface at the location of the planet.  Their results suggest that a planet with 
a mass $>~9~M_J$ at $\approx$~20 AU would create a depletion 
in sub-$\mu$m surface grains that would be detectable in our PSF-subtracted 
1400\AA, 1650\AA, and 3300\AA\ data.
However, a depletion is not detected in our UV or optical observations, and its
absence places an upper limit on the mass of a planet in the disk of GM~Aur to
a mass of $<~9~M_J$, likely even lower, though the  model grid is too course to be more specific. 
Furthermore, the models by \citet{Zhu_12} and \citet{de_Juan_Ovelar_13} 
also place a $1~M_J$ lower limit on planet mass due to the detection of a cavity in mm and sub-mm 
data \citep{Calvet_05,Hughes_09,Grafe_11,Andrews_11}.  

\subsection{A Possible Molecular Outflow}
\label{subsec-possmolecularoutflow}

Molecular outflows launched from the inner disk are expected in very young star
$\plus$ disk systems \citep{Ercolano_09}, but their persistence in older systems is
less well explored. Extended H$_2$ emission has been detected around a host of young
objects, for example RU~Lupi  \citep[different signals in different apertures,][]{Herczeg_05a}, plus
T~Tau \citep{Walter_03,Saucedo_03} and
DG~Tau \citep{Schneider_13}.  Spectroscopically, it is clear that the
H$_2$ is coming from an emitting region that is not from the star but rather from the
inner part of the disk, thus it is extended spatially. For DG~Tau, the UV H$_2$ emission appears
as a limb brightened ``bubble" with a length of about 0$\farcs$3= 42~AU located toward
the approaching lobe of the outflow. A very similar morphology is observed in
the near-IR \citep{Agra-Amboage_14}.
The H$_2$ emission lines in several systems are also
red/blue-shifted, which is expected in the case of molecular outflows
\citep{Herczeg_05,Herczeg_06,France_12}; \citet{Herczeg_06} discuss the spectral evidence
for  other potential sources of H$_2$ emission as well.
\citet{France_12} found no clear evidence 
for extended UV fluorescent H$_2$ emission
lines for GM Aur in their Cosmic Origins Spectrograph (COS) data around 1450 \AA . The COS line
profiles they present do not show obvious 
extended wings, though there may be a hint of marginal 
additional flux at -30 to -40 km~s$^{-1}$ in their Fig.~3.  Otherwise, 
they are centered at the systemic velocity
and appear compatible with emission from the inner disk (at radius 0.5~AU).  Their analysis was designed 
to map the distribution of H$_2$ in the disk, assuming Keplerian motion of the fluorescent gas.

As discussed in \S \ref{FUV_structure}, we detect an extended signal aligned along  
the disk's semi-minor axis in the F140LP and F165LP data that we do not detect in any other bandpass. 
This structure projects beyond the  boundary where we expect to observe reflection and scattered 
light from the disk surface.  The protruding region of the feature is limb-brightened and 
its geometry appears cylindrical 
with a radius of 40 $\pm$ 10~AU.  A possible interpretation is that it is a gaseous
photoevaporative wind of H$_2$, with some of the smallest sub-$\mu$m dust grains embedded within, 
being driven from the disk surface just beyond the location of the 24~AU sub-mm cavity.  
We only detect the feature in the observations where numerous H$_2$ transitions exist 
(in the FUV), 
indicating its composition could be primarily molecular hydrogen.  The smallest dust grains are 
the most tightly coupled to the gas, so without detection in the F330W data, any dust grains
embedded within the gas must be either small in radius or low in abundance.   
The structure, being aligned with the system minor axis, may offer
an explanation for the 5\arcsec\ distant ``polar lobes'' seen in the NICMOS data
\citep{Schneider_03}, which they interpreted as likely due to shocked line emission
from a molecular outflow.  
We exercise some caution in this interpretation,
subject to confirmation of this feature, which would benefit from
more detailed modeling and further FUV spectroscopy specifically designed to
cover the candidate outflow region. 

The dominant excitation mechanisms in classical T~Tauri stars are the high-energy
X-ray and EUV photons that are produced during accretion.  These high-energy
photons photo-dissociate molecular gas and excite the atomic components that
remain, so they are easily detected in the optical forbidden emission lines
\ion{O}{1}, \ion{S}{2} etc. \citep{Schneider_13}. Strong UV \ion{C}{4}, \ion{He}{2}, 
\ion{Si}{4} and \ion{N}{5} 
lines have been detected towards GM~Aur 
\citep{Ardila_13}. Furthermore, \citet[their figure 2]{Schneider_13} show the bandpass of the
F140LP and F165LP filters superposed on the spectrum of the active T~Tauri
star DG Tau. \ion{C}{4}, \ion{He}{2}, and \ion{Si}{4} emission may also 
contribute to the flux in the 
F140LP filter. 
The \ion{C}{4} lines are the strongest features longward of 1300~\AA\ but as noted
in \citet{France_12}, the integrated H$_2$ flux is twice that in the \ion{C}{4}
doublet.
However, we have not detected any
extended atomic or ionic emission lines in the broadband optical WFPC2 imagery or
UV/optical STIS spectra, and 
conclude that it is not an atomic or ionic jet that is
responsible for the extended emission detected in the FUV ACS/SBC data.  
For clarification, we follow the terminology  convention of \citet{Klaassen_13,Ray_07}
\&\ \citet{Reipurth_Bally_01}, and  define a jet as high velocity gas
(>~100~km~s$^{-1}$) launched from the inner 0.1 AU of the disk, collimated by the
magnetic field. The geometry of the outflow is inconsistent with typical collimated
atomic jets. The outflow appears to launch from the inner portions of the outer
disk rather than the central star, and smoothly extends 190~AU $\pm$ 35~AU until it
is no longer detectable above the background.  There is also no detection of
accompanying H I in the outflow, so it is likely that the material does not
dissociate, which precludes X-ray and EUV as the excitation mechanism.  
H$_2$ is far more likely this far out from the
star rather than hot (up to $10^5$~K) gas.  The hot transition region lines
(\ion{C}{4}, \ion{Si}{4}) seem to be emitted from the stellar surface, while the 
molecular lines are circumstellar. These are further arguments against 
the possibility that the source of the
emission is hot. It is
possible that the H$_2$ is being excited by FUV photons at Ly$\alpha$.  If it
is fluorescently excited by Lyman $\alpha$ from the vicinity of the star, we should
expect the radial brightness of the structure to decline as $\approx r^{-2}$. 
However, we find that the  radial profile of the structure follows an $r^{-0.7 \pm
0.1}$ dependance.  One possibility this implies is that some fraction of the
fluorescent excitation could be shocked gas \citep{Herczeg_02}.  

GM~Aur may possibly present the first direct detection of the dominant mechanism
responsible for clearing gas from the outer disk.  If this phenomenon were
representative of other transitional disks at 1-2 Myr, then it would place tight
constraints on the time frame over which gas giants may form.  How common this 
could be among transitional disks and over what portion of a disk's lifetime this 
continues are points that can only be addressed with observations of additional 
transitional disks in the FUV.

\subsection{Summary of Results}

$\bullet$ \ We confirm previous results for disk orientation and inclination, but at 
shorter wavelengths than those in the literature.  Additionally, we have combined multiple datasets 
from the literature and unpublished data to create a detailed SED that spans nearly 4 orders
of magnitude (from 0.14 $\mu$m to 860 $\mu$m).

$\bullet$ \ We resolve the disk down to a radius corresponding to 15 AU (the distance to GM~Aur 
assumed to be 140 pc) in optical and FUV wavelengths.  We do not detect a change in the 
radial surface brightness profile at or near the location of the sub-mm cavity wall.  We conclude
that small grain dust and gas exist within the cavity, which is consistent with models that describe dust filtration
via planet-disk interaction \citep{Zhu_12,Dong_12a,de_Juan_Ovelar_13}.

$\bullet$ \ Comparing the surface brightness of the disk imaged at the multiple wavelengths discussed 
here and reported in \citet{Schneider_03} with grain models (see \citet{Whitney_92,Kim_94,Cotera_01,Wood_02}), 
we conclude that the surface of the disk is populated by small grains.  

$\bullet$ \ The FUV observations detect a signal, undetected at longer wavelengths, 
that extends along the disk semi-minor axis.  One possible explanation we put forth is that it is a FUV
photoevaporative disk wind composed of H$_2$ and small grain dust.  However, radial velocity 
measurements along with additional FUV long-slit spectral data are needed in the future, in order to test this hypothesis.

\subsection{Implications for the Future}

As mentioned in \S \ref{subsec-possmolecularoutflow}, small scale H$_2$ 
molecular outflows have been detected in T~Tauri stars, especially in 
the near-IR \citep{Beck_08,Beck_12}, but also 
in the FUV \citep[DG Tau;][]{Schneider_13}. They could be something other
than thermally driven winds.  For example,
at $r=40$~AU around a $1.2~M_{\odot}$ star the escape velocity would be 
$v_{esc}\sim 7$~km~s$^{-1}$, which
would require a gas temperature $T_{gas}>5000$~K, high for the molecular gas
component.
GM~Aur allows possibly the first direct imaging detection of such a molecular outflow 
from a T-Tauri star in the FUV. If 
confirmed in future observations of GM~Aur, its presence would have
far-reaching implications.  Observations with ALMA could determine 
the abundance \&\ chemical composition of 
GM~Aur's disk, as well as 
provide high precision radial velocity data for the gas in the
extended region \citep{Bruderer_14,Klaassen_13,Mathews_13}.  
In the past, a topic of much speculation has been 
over what time frame and by what mechanism gas is
cleared from the outer disk.  H$_2$ is the
dominant gas species in disks.  Therefore, if this outflow feature is composed primarily 
of H$_2$, and found in future observations to be present at an early age for a significant 
number of star + disk systems, it will constrain the time frame over which gas giants may form.  
ALMA data could also have an impact on how we approximate gas to dust ratios in Monte Carlo radiative 
transfer and hydrodynamical models of transitional disk systems \citep{Bruderer_14}. 

When available, high-contrast imagery in FUV and short-wavelength 
optical bandpasses enhances our ability to determine whether small-grain dust
exists within the cavities of transitional disks.  Transitional disks with larger 
cavities than GM~Aur would constrain the dust
opacity and particle size distribution, as well as place limits on the ice content at the dust
disk surface.  Higher S/N data, such as might be provided by the next generation of
UV instrumentation,  are needed to probe ice chemistry in the outer dust disk surface.
They could also more stringently constrain the abundance of pure ice grains from the most
abundant ice species. Converting flux data into mass loss rates - which can then be
quantitatively compared with predicted photo-evaporation and photo-dissociation rates
- requires velocity data at the location of the molecular H$_2$ outflow.  The loss
rate of the dominant gas species in the disk is directly related to the time allowed
for gas giant formation.  It is important for transitional disk investigators to
fully explore the feasibility of obtaining high-contrast FUV imagery on these objects
while the opportunity exists, as there is no approved successor to HST with a FUV
imaging capability.

\section{ACKNOWLEDGEMENTS}

This work is, in part, based on observations made with the NASA/ESA Hubble Space
Telescope, obtained at the Space Telescope Science Institute, which is operated by
the Association of Universities for Research in Astronomy, Inc., under NASA contract
NAS 5-26555. JBH was supported in part by funding from the NASA Kentucky Space
Grant Consortium, Award \# 3049024102-11-175.  Data used in this study were obtained
under programs HST-GO-10864, HST-GO-11336, and HST-GO-12016.  We thank A.M. Hughes
and S. Andrews for the SMA and Plateau de Bure data.
AB was supported by the
grant HST-GO-11336.01-A, for which observing time was granted by the Chandra X-ray
Observatory peer review.    The authors thank the support staff members of the IRTF
telescope for assistance in obtaining the SED data, and the IR\& D program at The
Aerospace Corporation.  We also acknowledge support from NASA NNH06CC28C (M.L.S.) and
NNX09AC73G (CAG. and MLS.). We would like to thank Kenneth Wood and Michael J. Wolff 
for their quick response to our questions about their dust grain models 
during private communications. Finally, we thank two anonymous referees for many
suggestions which significantly improved this paper.
We dedicate this paper to the memory of Bruce Woodgate,
a colleague, mentor and friend who died during the preparation of this paper.

{\it Facility:} \facility{HST (NICMOS, STIS, ACS)}

\clearpage

\bibliographystyle{apj}

\bibliography{MasterBib_JHornbeck_bu20160429}

\vfill\eject

\begin{deluxetable}{lccl}
\tablecaption{HST Filter Characteristics}
\tablehead{
        \colhead{Filter} & 
        \colhead{wavelengths (\AA )} & 
        \colhead{plate scale} & 
        \colhead{angular resolution} \\
        \colhead{} & 
        \colhead{} & 
        \colhead{} & 
        \colhead{80\% encircled energy}               
	} 
\startdata
F140LP & 1360-1580 50\% max transmission   & $0.034\times 0.030''$ & 0.30$''$ (based on F150LP) \\
F165LP & 1640-1830 50\% max transmission  & $0.034\times 0.030''$ & 0.30$''$ (based on F150LP) \\
F330W  & $\lambda_{eff}=3376$, $\Delta \lambda=529$  & $0.028\times 0.025''$ & 0.25$''$ (based on F220W) \\
F555W  & $\lambda_{pivot}=5439$, $\Delta \lambda=1236$  & $0.046\times 0.046''$ & 0.15$''$  \\
\enddata
\label{tab-centralwavelengths}
\end{deluxetable}

\begin{deluxetable}{ccrcccr}
\tablewidth{0pt}
\tabletypesize{\small}
\tablecaption{HST Observations}
\tablehead{
        \colhead{Object}     & 
        \colhead{Dataset}    & 
        \colhead{Program ID} & 
        \colhead{Date}       & 
        \colhead{Instrument} & 
        \colhead{Filter}     & 
        \colhead{Exp (s)} \\
	}
\startdata
GM~Aur    & JA5M01020-30 & 11336 & 2008-08-13 & ACS/SBC     & F140LP & 2520,2552 \\ 
%GM~Aur    & JA5M01020 & 11336 & 2008-08-13 & ACS/SBC     & F140LP & 2520 \\ 
GM~Aur    & JA5M01010 & 11336 & 2008-08-13 & ACS/SBC     & F165LP & 2528 \\ 
GM~Aur    & J8MS09PNQ & 9812  & 2003-12-31 &  ACS/HRC    & F330W  & 360 \\
GM~Aur    & U2RD0401T-04T & 6223  & 1995-07-29 & WFPC2/PC1   & F555W  & $4\times 30$ \\

\cutinhead{PSF Template Observations}  \\ 
NQ~UMa    & JBDF07010 & 12016 & 2010-06-05 & ACS/SBC     & F140LP & 2648\phantom{.6} \\
NQ~UMa    & JBDF06010 & 12016 & 2010-06-03 & ACS/SBC     & F165LP & 2648\phantom{.6} \\
HD~202560 & J8HV03031 & 9655  & 2002-12-13 & ACS/HRC     & F330W  & $2\times 2$\phantom{.6} \\
CY~Tau    & J8MSA1MUQ & 9812  & 2003-12-27 & ACS/HRC     & F330W  & 360\phantom{.6} \\
DS~Tau    & J8MS08HEQ & 9812  & 2003-12-30 & ACS/HRC     & F330W  & 360\phantom{.6} \\
HD~283572 & U2RD0301T-04T & 6223  & 1995-07-23 & WFPC2/PC1   & F555W  & $4\times 1.6$ \\
\cutinhead{STIS Observations}  \\
GM~Aur & OB3R04050 & 11608 & 2011-09-11 & STIS & G140L & 3020 \\
GM~Aur & OB3R04040 & 11608 & 2011-09-11 & STIS & G230L & 1231 \\
GM~Aur & OB3R04010 & 11608 & 2011-09-11 & STIS & G430L & 35 \\
GM~Aur & OB3R04020 & 11608 & 2011-09-11 & STIS & G750L & 8 \\

\enddata
\label{tab-observations}
\end{deluxetable}

\begin{deluxetable}{lccclcl}
\tablewidth{0pt}
\tabletypesize{\small}
\tablecaption{Scattered Light Properties}
\tablehead{
        \colhead{$\lambda$}                                                     & 
        \colhead{IWA (AU)}                                                      & 
        \colhead{Extent\tablenotemark{\dag}}            & 
        \colhead{Aspect Ratio}                                          & 
        \colhead{Inclination\tablenotemark{\ddag}}      &
        \colhead{Aperture radius}                                       & 
        \colhead{$F_{disk}/F_{star}$}                                            
}
\startdata
F140LP & 0$\farcs$11 (15) & \ldots                                                      & \ldots        & \ldots                                        & 3$\farcs$4    & 0.321 $\pm$ 0.12      \\  
F165LP & 0$\farcs$11 (15) & \ldots                                                      & \ldots        & \ldots                                        & 3$\farcs$4    & 0.310 $\pm$ 0.08      \\ 
F330W  & 0$\farcs$10 (14) & 3$\farcs$13 $\pm$ 0$\farcs$13       & 0.54          & 57\degree\ $\pm$ 9\degree & 3$\farcs$4        & 0.130 $\pm$ 0.03      \\  
F555W  & 0$\farcs$15 (28) & 4$\farcs$16 $\pm$ 0$\farcs$10       & 0.59          & 53\degree\ $\pm$ 3\degree & 3$\farcs$4        & 0.129 $\pm$ 0.03      \\ 
\enddata
\tablenotetext{\dag}{Full extent of the disk measured along the disk major axis; 60\degree\ $\pm$ 3\degree\ east of north.}
\tablenotetext{\ddag}{Inclination calculated from extent and aspect ratio}
\tablecomments{The disk of GM~Aur in the F140LP and F165LP data has features that do not allow the
measurement of the extent and aspect ratio of the disk  (Fig. \ref{fig-psfsubtractions}). The disk
scattering fraction, $F_{disk}/F_{star}$, is the ratio comparing the light detected from the disk
between 0$\farcs$30 and 3$\farcs$4, after PSF subtraction of the light coming from the star + disk
system prior to PSF subtraction.   We choose the region exterior to 0$\farcs$30, rather than the IWA
of each dataset, so that we can compare our disk scattering fraction results to the NICMOS results
presented in \citet{Schneider_03}.}
\label{tab-scatteredlight}
\end{deluxetable}

\begin{deluxetable}{lccccccc}
\tablecaption{Power Law Indices for HST Optical Images}
\tablehead{
        \colhead{Filter}                                & 
        \colhead{rotation offset}               & 
        \multicolumn{3}{c}{east side}   & 
        \multicolumn{3}{c}{west side}   \\
                                                                &              
                                                                        & 
        $\gamma$                                        & 
        reduced $\chi^2$                                & 
        prob                                                    & 
        $\gamma$                                        & 
        reduced $\chi^2$                                & 
        prob                                                    } 
\startdata
F330W & -5$^\circ$           & $1.54\pm 0.02$ & 2.89    & 0.00  & $2.01\pm0.01$     & 4.37      & 0.00 \\
F330W & \phantom{-}0$^\circ$     & $1.41\pm 0.03$ & 0.89   & 0.65   & $2.17\pm0.04$     & 1.02  & 0.43 \\
F330W & +5$^\circ$           & $1.43\pm 0.02$ & 1.99    & 0.00  & $1.97\pm0.02$     & 1.90  & 0.00 \\ \hline
F555W & -5$^\circ$           & $1.67\pm 0.06$ & 2.25    & 0.01  & $2.02\pm0.06$     & 0.19  & 0.99 \\
F555W & \phantom{-}0$^\circ$     & $1.63\pm 0.06$ & 1.54    & 0.12  & $2.06\pm0.05$     & 0.29  & 0.98 \\
F555W & +5$^\circ$           & $1.69\pm 0.06$ & 0.69    & 0.73  & $2.15\pm0.06$     & 0.13  & 0.99 \\
\enddata
\label{tab-f555w_surfbrightness}
\end{deluxetable}

%figure 1

%figure 2

%figure 3

%figure 4

%figure 5

%figure 6

%figure 7

%figure 8

%figure 9

%figure 10

%Figure 11

\end{document}